\documentclass[twocolumn,
  nobibnotes,
  aps,
  prb,
  showpacs,
  amsmath,
  amssymb,
  superscriptaddress,
  floatfix,
  10pt]{revtex4-1}

\usepackage{graphicx}
\usepackage{bm}
\usepackage{amsmath}

\usepackage{color}

\usepackage[T1]{fontenc}
\usepackage{bbold}

\begin{document}

\title{Transversal magnetoresistance in Weyl semimetals}%

\author{J. Klier}

\affiliation{Institut f\"ur Nanotechnologie, Karlsruhe Institute of Technology, 76021 Karlsruhe, Germany}

\author{I.V. Gornyi}

\affiliation{Institut f\"ur Nanotechnologie, Karlsruhe Institute of Technology, 76021 Karlsruhe, Germany}

\affiliation{\mbox{Institut f\"ur Theorie der Kondensierten Materie, Karlsruhe Institute of Technology, 76128 Karlsruhe, Germany}}

\affiliation{A. F. Ioffe Physico-Technical Institute, 194021 St.~Petersburg, Russia}

\author{A.D. Mirlin}

\affiliation{Institut f\"ur Nanotechnologie, Karlsruhe Institute of Technology, 76021 Karlsruhe, Germany}

\affiliation{\mbox{Institut f\"ur Theorie der Kondensierten Materie, Karlsruhe Institute of Technology, 76128 Karlsruhe, Germany}}

\affiliation{Petersburg Nuclear Physics Institute, 188350 St.~Petersburg, Russia}

\date{\today}%
\begin{abstract}

We explore theoretically the magnetoresistivity of three-dimensional
Weyl and Dirac semimetals in transversal magnetic fields
within two alternative models of disorder:
(i) short-range impurities and (ii) charged (Coulomb) impurities.
Impurity scattering is treated using the self-consistent Born approximation.
We find that an unusual broadening of Landau levels leads to a variety of
regimes of the resistivity scaling in the temperature--magnetic field plane.
In particular, the magnetoresistance is non-monotonous for the white-noise disorder model.
For $H\to 0$ the magnetoresistance for short-range impurities
vanishes in a non-analytic way as $H^{1/3}$.
In the limits of strongest magnetic fields $H$, the magnetoresistivity
vanishes as $1/H$ for pointlike impurities, while it is linear and positive
in the model with Coulomb impurities.

\end{abstract}
\maketitle

\section{Introduction}

Topological materials and structures represent one of the central research directions in the modern condensed matter physics. One of the classes of such materials is topological insulators and superconductors which possess a bulk gap and topologically protected surface excitations with massless Dirac spectra. Another important class is gapless materials with topologically protected Fermi points or nodal lines. The most well-known example is graphene whose dispersion is characterized by two Fermi points where the valence and the conduction bands touch. Excitations in the vicinity of these points have a linear dispersion and can be viewed as two-dimensional Dirac fermions.

Three-dimensional counterparts of graphene are Dirac semimetals. In such a material the dispersion near the nodal point is characterized by $4\times4$ Dirac Hamiltonian, i.e. the conductance and the valence bands have an additional twofold degeneracy. Experimental realizations of Dirac semimetals include  Cd$_{3}$As$_{2}$ \cite{RIS_0} and Na$_{3}$Bi \cite{Liu21022014}; further candidate materials have been recently discussed  \cite{2014arXiv1411.0005G}.  The twofold degeneracy discussed above can be lifted if either spatial inversion or time-reversal symmetry is broken. The four-component Dirac solution then decouples into two independent two-component solutions representing two Weyl fermions of opposite chirality. Thus, each of the Dirac points then splits into two Weyl points without any additional degeneracies. Recent experiments provided evidence that TaAs \cite{2015arXiv150204684L,2015arXiv150203807X}
 and NbAs\cite{2015arXiv150401350X} can be classified as Weyl semimetals \cite{Xu16012015,2015arXiv150700521Y}.
Further promising candidates for Weyl semimetals include pyrochlore iridates \cite{PhysRevB.83.205101}, topological insulator heterostructures \cite{PhysRevLett.107.127205}, and Cd$_{3}$As$_{2}$ with lowered symmetry \cite{PhysRevB.88.125427}.
In the rest of the paper we will use the term ``Weyl semimetal'' in a broader sense, including also the degenerate case of Dirac semimetals.

Transport properties of Weyl semimetals are highly peculiar (for various theoretical aspects of the problem, see, e.g., Refs. \onlinecite{PhysRevB.84.235126,PhysRevLett.108.046602,reviewVish,0953-8984-27-11-113201,PhysRevB.89.054202,
PhysRevLett.113.026602,PhysRevB.88.104412,PhysRevB.89.014205, PhysRevB.91.035133, PhysRevLett.114.257201,PhysRevLett.113.247203, PhysRevB.91.245157, PhysRevLett.114.166601, 2015arXiv150507374S,PhysRevB.91.035202,PhysRevB.91.195107,PhysRevB.83.205101} and references therein).
In particular, weak disorder (with a strength below a certain critical value) has a negligible effect on the density of states. Specifically, the density of states vanishes quadratically in energy around the Weyl point despite the presence of disorder. Furthermore, the limits $T\to 0$ and $\omega\to 0$  (where $\omega$ is the frequency) are not interchangeable for the behavior of the conductivity (assuming that the chemical potential is at the Weyl point). While sending frequency to zero first results in a finite conductivity, the zero-$T$ {\it ac} conductivity vanishes \cite{PhysRevB.83.205101,PhysRevB.84.235126,PhysRevLett.108.046602,PhysRevB.89.245121} in the limit $\omega\to 0$ as $|\omega|$. In the strong-disorder regime these singularities are eliminated.

Behavior of Weyl semimetals in the external magnetic field is also expected to be very nontrivial. This is related, first of all, to the unconventional Landau quantization of Dirac fermions. Furthermore, since a single Weyl node displays a chiral anomaly, a possibility to control the valley polarization as well as a large anomalous Hall effect are expected.~\cite{reviewVish,0953-8984-27-11-113201,PhysRevLett.111.027201} Much attention has been recently put on the longitudinal
magnetoresistance in Weyl semimetals that originates from the chiral anomaly.~\cite{PhysRevB.88.104412, PhysRevB.89.085126, PhysRevLett.113.247203, PhysRevB.91.245157, 2015arXiv150302069G, 0953-8984-27-15-152201, 2015arXiv150606577S, RIS_5}

In this paper, we develop a theory of the transversal magnetoresistivity of a Weyl semimetal.
We are particularly interested in the range of sufficiently strong magnetic fields,
such that the Landau quantization is important.
\footnote{Recently, a mechanism of linear magnetoresistance due to recombination in compensated metals at non-quantized
magnetic field has been proposed for finite-geometry samples in Ref.~\onlinecite{PhysRevLett.114.156601}. In the present work, we
 focus on infinite systems.}
One of the motivations for our work was a spectacular experimental observation of a large,
approximately linear magnetoresistance in the Dirac semimetals Cd$_{3}$As$_{2}$ and TlBiSSe
in strong magnetic field.~\cite{RIS_1, 2014arXiv1405.6611F, 0953-8984-27-15-152201, Novak2015}
Quantum linear magnetoresistance has been obtained by Abrikosov in a seminal paper, Ref.~\onlinecite{PhysRevB.58.2788},
for Dirac semimetals in the extreme limit  when only one Landau level is filled (i.e., the cyclotron frequency exceeds the temperature).
The linear behavior was traced back to the magnetic-field dependent screening of charged (Coulomb) impurities.

We consider a general case of arbitrary relation between the magnetic field and temperature such that,
depending on the regime, the magnetoresistance is dominated by contributions from the zeroth Landau level,
other separated Landau levels, and overlapping Landau levels. We address
two alternative models of disorder: (i) short-range impurities and (ii) Coulomb impurities.
We show that for pointlike impurities the transition between the weak-disorder and the strong-disorder phases persists
in the presence of magnetic field and explore singularities of the weak-disorder phase.
Further, we find that an unusual broadening of Landau levels leads to a variety of regimes of
the resistivity scaling in the temperature-magnetic field plane.
The transversal magnetoresitance is found to be non-monotonous for the model of weak white-noise disorder.
Remarkably, in the limit of $H\to 0$, the magnetoresistance for short-range impurities
shows a non-analytical behavior $H^{1/3}$. For strong pointlike impurities and for charged impurities
we find a positive quadratic magnetoresistance at $H\to 0$. In the limit of strongest magnetic fields $H$,
the magnetoresistivity vanishes as $1/H$ for pointlike impurities,
while it is linear and positive in the model with Coulomb impurities, in agreement with Abrikosov's result~\cite{PhysRevB.58.2788}
and recent experimental findings.

The paper is organized as follows. In Sec.~\ref{sec:BA}, we outline the implementation of the
Born approximation in the context of Weyl semimetals subjected to a magnetic field in the presence
of pointlike impurities. We find that, in analogy with the zero-magnetic-field case,
one has to distinguish between weak and strong disorder regimes separated by a phase transition.
In Sec.~\ref{SCBA} we develop the formalism of the self-consistent Born approximation (SCBA) and
analyze the broadening of Landau levels due to disorder and magnetic field.
Section \ref{sec:conductivity} presents the formalism for calculation of the conductivity
in magnetic field for the model of white-noise disorder.
In Sec.~\ref{sec:CoulombImp}, we extend our analysis to the case of charged impurities.
In Sec. \ref{sec:magnetoresistance} we use the obtained results to calculate and analyze
the magnetoresistance for both models of disorder.
Our findings are summarized in Sec.~\ref{sec:summary}.

\section{Landau-levels spectrum in Weyl semimetals}
\label{sec:BA}

\subsection{Clean case}
\label{sub:clean}
In the presence of a constant homogeneous magnetic field $H$ in $z$ direction, electrons of a single Weyl node are described by the Weyl Hamiltonian
\begin{equation}
\mathcal{H}\left(\textbf{p}\right)=\int d^{3}\textbf{r}\Psi^{\dagger}(\textbf{r})v\bm{\sigma}\left(\textbf{p}-\frac{e}{c}\textbf{A}\right)\Psi(\textbf{r}),
\end{equation}
where  $\boldsymbol{\sigma}$ denote the Pauli matrices and  $\textbf{A}(\textbf{r})=(0, Hx, 0)$ is the vector potential. (We have chosen the Landau gauge.)
The eigenfunctions of this Hamiltonian have two components ($\alpha,\beta=1,2$) in the space spanned by  $\boldsymbol{\sigma}$.
Positions of the Landau levels (LLs) in a clean Weyl semimetal are given by
\begin{align}\label{eigst}
\varepsilon_{0}=& vp_{z},\\
\varepsilon_{n}^{(\pm)}=&\pm v\sqrt{p_{z}^{2}+\frac{2n}{l_{H}^2}},
\end{align}
where $l_H=(eH/c)^{-1/2}$ is the magnetic length and $\Omega=v\sqrt{2eH/c}$ is the
distance between the zeroth and first LL. We set $\hbar=1$ throughout the paper.
In the following, we choose the energy bands such~\cite{PhysRevB.58.2788} that the wave
function of the clean zeroth LL has only component 1 in the pseudospin space:
\begin{align}\label{eigfunc}
\Psi_{n1}^{(\pm)}(\textbf{r})&=\frac{1}{\sqrt{2}}\left(1+\frac{vp_{z}}{\varepsilon_{n}^{\pm}}\right)^{1/2}\!
\frac{e^{i(p_{y}y+p_{z}z)}}{L} \phi_{n}(x-l^{2}p_{y}),
\nonumber\\
\Psi_{n2}^{(\pm)}(\textbf{r})&=\mp\frac{i}{\sqrt{2}}\left(1-\frac{vp_{z}}{\varepsilon_{n}^{\pm}}\right)^{1/2}\!\frac{e^{i(p_{y}y+p_{z}z)}}{L}\phi_{n-1}(x-l^{2}p_{y})
\end{align}
for $n > 0$
and
$\Psi_{01}^{(\pm)}=\theta(\pm p_z)\phi_0, \ \Psi_{02}^{(\pm)}=0$ for $n=0$.
Here $\phi_n$ are the normalized eigenfunctions of free electrons in magnetic field and $\theta$ denotes the Heaviside step function.
The retarded bare Green function $\hat{G}_0$ of the clean system is conveniently represented as a matrix in the pseudospin space of bands
$\alpha,\beta = 1,2$:
\begin{equation}
G^{(0)}_{\alpha\beta}=\sum_{n\geq 0,\lambda=\pm}\frac{\Psi_{n\alpha}^{(\lambda)}{\Psi_{n\beta}^{(\lambda)}}^*}{\varepsilon+i0-\varepsilon_{n}^{\lambda}}.
\label{bareGF}
\end{equation}
It is worth noting that the summation of $\lambda$ eliminates the theta-functions $\theta(\pm p_z)$ in the $n=0$ term in Eq.~\eqref{bareGF},
so that the integration over $p_z$ in what follows will always be performed from $-\infty$ to $\infty$.

\begin{figure}
\begin{center}
\includegraphics[scale=0.7]{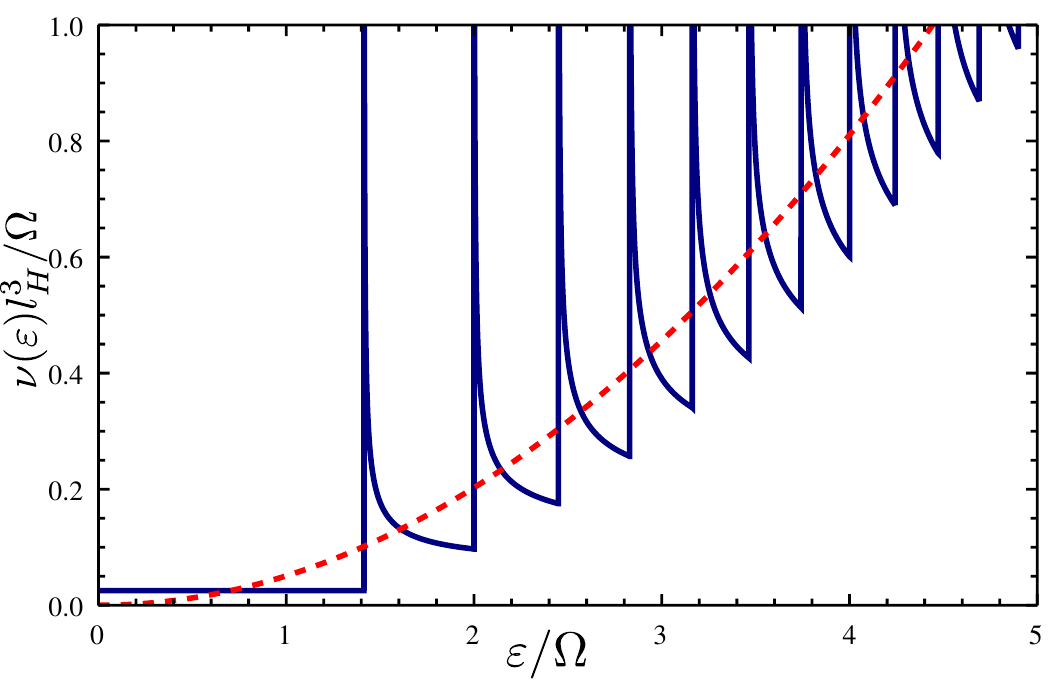}
\caption{Density of states of a clean Weyl semimetal. The blue solid line shows the DoS in a finite magnetic field, Eq.~(\ref{densityofstates}),
while the red dashed line corresponds to the zero magnetic field, Eq.~(\ref{B0DOS}).}
\label{density}
\end{center}
\end{figure}

In the zero magnetic field the clean density of states (DoS) behaves quadratically in energy:
\begin{equation}
\nu(\varepsilon)=\frac{\varepsilon^2}{2\pi^{2}v^{3}}.
\label{B0DOS}
\end{equation}
In the presence of magnetic field, the DoS acquires a sawtooth form with square root singularities originating from the one-dimensional ($p_z$) dispersion of each Landau band (cf. Ref.~\onlinecite{density})
\begin{equation}\label{densityofstates}
\nu(\varepsilon)=\frac{1}{4\pi^{2}l^{2}_H v}\left[1+2\sum_{n=1}^{\varepsilon^{2}l^{2}_H/2}\frac{|\varepsilon|}{\sqrt{\varepsilon^{2}-2nv^2/l^{2}_H}}\right].
\end{equation}
The DoS is visualized in Fig.~\ref{density}.

\subsection{Introducing disorder}
\label{sub:disorder}
We consider now the effect of disorder. The impurity scattering generates a self-energy $\hat{\Sigma}(\textbf{p}, \varepsilon)$ in the (impurity-averaged) Green function,
\begin{equation}\label{green}
\hat{G}(\textbf{p}, \varepsilon)=\left\langle\frac{1}{\varepsilon-\mathcal{H}}\right\rangle
=\frac{1}{\varepsilon-v\boldsymbol{\sigma}\cdot\left(\textbf{p}-\frac{e}{c}\textbf{A}\right)-{\hat{\Sigma}}(\textbf{p}, \varepsilon)},
\end{equation}
which is a matrix in the pseudospin space (in which the Pauli matrices $\sigma$ operate).

We will assume that the disorder potential is diagonal in both spin and pseudospin indices and neglect scattering between different Weyl nodes.
We will discuss this approximation and its limitations in the end of the paper, Sec.~\ref{sec:summary}.
Clearly, in the absence of internode scattering, the structure in the node space will be trivial for all quantities.
We do not show it explicitly below; the calculated density of states and the conductivities are those per Weyl node.

We will first consider a model of pointlike impurities and later analyze generalization to the case of Coulomb impurities.
The impurity potential has then the form
\begin{equation}
\hat{V}_{\text{dis}}(\textbf{r})=u_{0}\sum_{i}\delta(\textbf{r}-\textbf{r}_{i})
\mathbb{1},
\end{equation}
where $\mathbb{1}$ is the unit matrix in the pseudospin space.
In view of the matrix structure of the impurity potential $\hat{V}_{\text{dis}}(\textbf{r})$,
the impurity correlator $\hat{W}$ becomes a rank-four tensor. The self-energy reads
\begin{equation}\label{scba}
\Sigma_{\alpha\beta}(\textbf{r},\textbf{r}')=
\int\frac{d^{3}q}{(2\pi)^{3}}W_{\alpha\gamma\beta\delta}(\textbf{q})
e^{i\textbf{q}\cdot(\textbf{r}-\textbf{r}')}G_{\gamma\delta}(\textbf{r},\textbf{r}').
\end{equation}
For a diagonal impurity potential, the impurity correlator is diagonal as well, which is expressed as
\begin{equation}\label{correlator}
W_{\alpha\gamma\beta\delta}(\textbf{q})=\gamma
\delta_{\alpha\gamma}\delta_{\beta\delta},
\end{equation}
where $\gamma=n_\text{imp}u_0^2$.
The self-energy is diagonal in the energy-band space.
However, in the presence of magnetic field, the self-energy is no longer proportional to the unit matrix:
\begin{equation}
\hat{\Sigma}=\text{diag}(\Sigma_{1},\Sigma_{2}).
\end{equation}
This asymmetry originates from the asymmetry of states in the zeroth LL.
In the clean case, the states of the zeroth LL are only present in one energy band.
Later we will see that a strong impurity scattering eliminates this asymmetry.

We switch to LL representation so that $\hat{G}=\hat{G}(\varepsilon,p_z,n)$ and
$\hat{\Sigma}=\hat{\Sigma}(\varepsilon,p_z,n)$. The diagonal components of the matrix Green function (\ref{green}) that determine the self-energy
read:
\begin{eqnarray}
G_{11}&=&\frac{\varepsilon-\Sigma_2+vp_z}{(\varepsilon-\Sigma_1-vp_z)(\varepsilon-\Sigma_2+vp_z)-\Omega^2 n},
\label{G11}
\\
G_{22}&=&\frac{\varepsilon-\Sigma_1-vp_z}{(\varepsilon-\Sigma_1-vp_z)(\varepsilon-\Sigma_2+vp_z)-\Omega^2 (n+1)}.\nonumber
\\
\label{G22}
\end{eqnarray}
In general, the self-energy depends on energy and on the LL index, $\hat{\Sigma}=\hat{\Sigma}(\varepsilon,p_z, n)$. However, for a white-noise disorder, the dependences on $n$ and $p_z$ drop out.

\subsection{Born approximation}
\label{subsec:BA}
We start with the Born approximation, where
we neglect the self-energies in Green's functions (\ref{G11}) and (\ref{G22})
for the calculation of self-energies:
\begin{eqnarray}
\Sigma_{1}^R(\varepsilon)&=&\frac{\gamma}{2\pi l_H^2}\!\sum_{n\geq0}
\int_{-\infty}^{\infty}\!\!\frac{dp_z}{2\pi}\frac{\varepsilon+vp_z}{(\varepsilon+i 0)^2-\Omega^2 n - v^2 p_z^2},
\label{Sigma1BA}
\\
\Sigma_{2}^R(\varepsilon)&=&\frac{\gamma}{2\pi l_H^2}\!\sum_{n\geq1}
\int_{-\infty}^{\infty}\!\!\frac{dp_z}{2\pi}\frac{\varepsilon-vp_z}{(\varepsilon+i 0)^2-\Omega^2 (n+1) - v^2 p_z^2}.
\nonumber
\\
\label{Sigma2BA}
\end{eqnarray}
The summation over $n$ here should, in fact, be restricted by an upper cut-off  $N_\text{max}$, as will be discussed below.
After shifting the summation over $n$ in $\Sigma_{2}$,
we see that the two self-energies differ only by the absence of the $n=0$ term in $\Sigma_{2}$:
\begin{equation}
\Sigma_1-\Sigma_2=\frac{\gamma}{2\pi l_H^2}
\int_{-\infty}^{\infty}\!\frac{dp_z}{2\pi}\frac{\varepsilon+vp_z}{(\varepsilon+i 0)^2 - v^2 p_z^2}
\simeq -i A+\frac{2 A \varepsilon}{\pi \Lambda},
\label{Sigma1-2BA}
\end{equation}
where, using $l_H^2=2(v/\Omega)^2$, we have introduced
\begin{equation}
A=\frac{\gamma \Omega^2}{8 \pi v^3},
\label{A}
\end{equation}
and $\Lambda$ is the bandwidth.
The necessity to introduce the ultraviolet cut-off $\Lambda$ originates from the
approximation of a true energy dispersion by the Dirac-fermion one, which is, in fact,
a low-energy approximation. Our analysis is applicable for $\varepsilon,\Sigma(\varepsilon) \ll \Lambda$.

In view of Eq.~(\ref{Sigma1-2BA}), it is sufficient to calculate $\Sigma_1$:
\begin{eqnarray}
\text{Im}\Sigma_1(\varepsilon)&=&-A |\varepsilon|\ \sum_{n=0}^{N_\varepsilon}\frac{1}{\sqrt{\varepsilon^2-\Omega^2 n}},
\label{ImSigmaBA}
\\
\text{Re}\Sigma_1(\varepsilon)&=&-A \varepsilon\ \sum_{n=N_\varepsilon+1}^{N_\text{max}}\frac{1}{\sqrt{\Omega^2 n-\varepsilon^2}}.
\label{ReSigmaBA}
\end{eqnarray}
The imaginary part of the Born self-energy is produced by the Landau levels
below $\varepsilon$, while the real part is due to the contribution of the
Landau levels above $\varepsilon$. In Eqs.~(\ref{ImSigmaBA}) and (\ref{ReSigmaBA})
\begin{equation}
N_\varepsilon=\left[\frac{\varepsilon^2}{\Omega^2}\right],
\label{Ne}
\end{equation}
is the number of the Landau level below energy $\varepsilon$,
and the symbol $[\ldots]$ denotes the integer part of a number.
For the evaluation of the sum in Eq.~(\ref{ReSigmaBA}), we have introduced the upper cutoff $N_\text{max}$ which
is determined by the ultraviolet energy cutoff $\Lambda$ in the following way:
$N_\text{max}=\Lambda^2 /\Omega^2$ ($N_\text{max}$ is the index of the highest Landau level within the bandwidth $\Lambda$).

The sum in $\text{Re}\Sigma_1$ is dominated
by the upper limit $N_\text{max}$:
$\text{Re}\Sigma_1 \sim \varepsilon(A/\Omega)N_\text{max}^{1/2}$.
Assuming no Landau quantization at the ultraviolet energies $\sim \Lambda$,
we use the zero-$H$ result
\begin{equation}
\text{Re}\Sigma_1(\varepsilon) \simeq - \frac{\beta}{2} \varepsilon,
\label{ReSigma-beta}
\end{equation}
with
\begin{equation}
\beta=\frac{\gamma \Lambda}{2 \pi^2 v^3}.
\label{beta}
\end{equation}
The parameter $\beta$ quantifies the strength of disorder. For sufficiently strong
disorder, the real part of Born self-energy becomes larger than $\varepsilon$,
which clearly signifies the insufficiency of the simple Born approximation.
As we will see in Sec.~\ref{subsec:strong} below, the self-consistent treatment of strong disorder yields
a dramatic change of the behavior of the density of states for strong disorder.

\begin{figure}[t]
\centerline{\includegraphics[width=7cm]{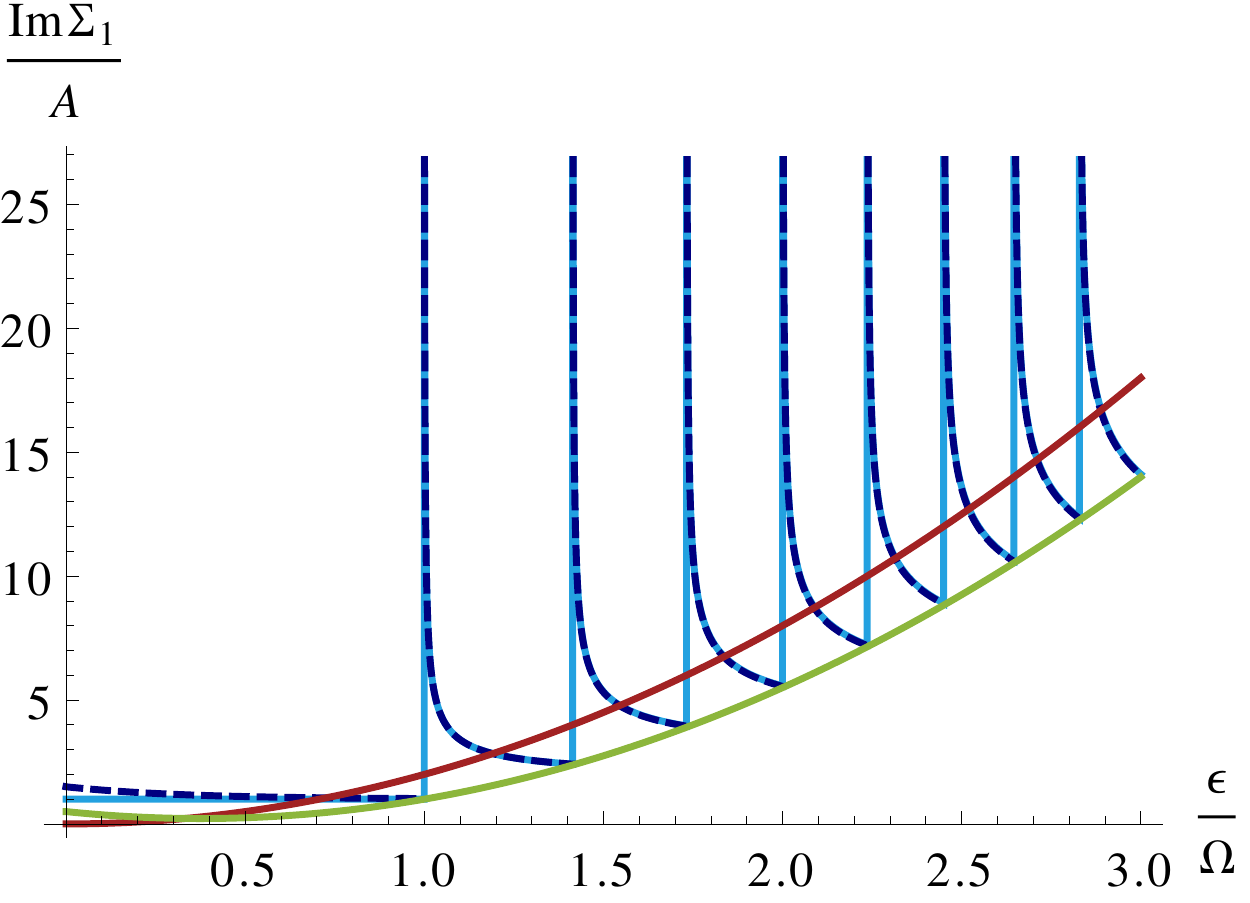}}
\caption{Disordered Weyl semimetal in Born approximation. $\text{Im}\Sigma_1(\varepsilon)$ in units of $A$.
Blue: Eq.~(\ref{ImSigmaBA}); Dark blue, dashed: Eq.~(\ref{ImSigmaBAEM});
Green: $2 (\varepsilon/\Omega)^2  - 3\varepsilon/2\Omega + 1/2$ [minima in Eq.~(\ref{ImSigmaBAEM})];
Red: $2(\varepsilon/\Omega)^2$ (result without magnetic field).}
\label{fig1}
\end{figure}

In what follows, however, we mostly focus on the limit of weak disorder, $\beta\ll 1$.
We absorb $\text{Re}\Sigma$ into the redefinition of the energy
$\varepsilon \to \tilde{\varepsilon}=\varepsilon(1+\beta/2)$, and neglect the difference
between $\tilde{\varepsilon}$ and $\varepsilon$.
For $|\varepsilon|<\Omega$, we find
\begin{equation}
\text{Im}\Sigma_1=-A, \qquad \text{Im}\Sigma_2=0.
\label{LLL:BA}
\end{equation}
For higher energies, using the Euler-Maclaurin formula for the sum over
$n<N_\varepsilon-1$, we express the imaginary part of the Born
self-energy as
\begin{eqnarray}
\text{Im}\Sigma_1(\varepsilon)
&\simeq &
-A\left[\frac{1}{\sqrt{\varepsilon^2-\Omega^2 N_\varepsilon}}-\frac{2|\varepsilon|}{\Omega^2}\sqrt{\varepsilon^2-(N_\varepsilon-1)\Omega^2}
\right.
\nonumber
\\
&+&
\left.
\frac{1}{2}\left(1+\frac{|\varepsilon|}{\sqrt{\varepsilon^2-(N_\varepsilon-1)\Omega^2}}\right)
+\frac{2\varepsilon^2}{\Omega^2} \right].
\label{ImSigmaBAEM}
\end{eqnarray}
This result is illustrated in Fig. \ref{fig1}.
The first term in the square brackets of Eq.~(\ref{ImSigmaBAEM}) is responsible for the square-root divergency at
the positions of LLs, whereas the last term yields the parabolic background similarly to the zero-$H$ case.

\section{Self-consistent Born approximation}
\label{SCBA}

We now turn to the self-consistent Born approximation (SCBA).
The self-consistent treatment is motivated by the presence of square-root singularities
in the Landau-level broadening (\ref{ImSigmaBAEM}) obtained within the Born approximation.
The introduction of disorder-induced self-energies in Green's functions should cut off
such divergencies.

The SCBA equations \eqref{scba}, \eqref{G11} and \eqref{G22} with the disorder correlator \eqref{correlator} acquires the form
(below $z=vp_z$):
\begin{eqnarray}
\Sigma_{1}(\varepsilon)&=&A \sum_{n\geq0}
\int_{-\infty}^{\infty}\!\!dz\ \frac{  \varepsilon-\Sigma_2+z}{(\varepsilon-\Sigma_1-z)(\varepsilon-\Sigma_2+z)-\Omega^2 n},
\nonumber
\\
\label{Sigma1SCBA}
\\
\Sigma_{2}(\varepsilon)&=&A \sum_{n\geq1}
\int_{-\infty}^{\infty}\!\! dz\ \frac{  \varepsilon-\Sigma_1-z}{(\varepsilon-\Sigma_1-z)(\varepsilon-\Sigma_2+z)-\Omega^2 n}.
\nonumber
\\
\label{Sigma2SCBA}
\end{eqnarray}
As above, we absorb the real parts of self-energies (determined by the ultraviolet cut-off $\Lambda$)
into the shifts of energies $\varepsilon \to \tilde{\varepsilon}$.
The density of states is given by
\begin{equation}
\rho(\varepsilon)=-\frac{1}{\pi} \text{Tr}\ \text{Im}G = -\frac{1}{\pi \gamma} \left(\text{Im}\Sigma_{1}+\text{Im}\Sigma_{2}\right).
\label{DOS}
\end{equation}

In the regime of well separated LLs (the corresponding conditions will be analyzed below),
the sum over the Landau levels is evaluated as follows.
Let us assume that the energy is close to the bottom of the $N$-th Landau level $N_\varepsilon\simeq N$.
Then the main contribution of the sum  over $n$ comes from the term $n=N$. We thus single out the term of $n=N$ from the sum
and evaluate it separately from the sum over the remaining Landau levels.

We note that the self-consistent treatment of the LL broadening is fully justified for weak disorder
and $\varepsilon\ll \Omega$. All the renormalization effects not captured by the SCBA affect the real part of the
self-energy, see discussion in Ref.~\onlinecite{PhysRevB.77.195430} where the SCBA was employed for 2D Dirac fermions in graphene.
In that case, disorder was marginally relevant and its effect could be incorporated through
the renormalization of parameters (induced by the contributions of higher LLs)
that enter the SCBA equations for a given Landau level.
In the present 3D case, the renormalization of $\text{Re}\Sigma$ can be neglected for the case of
weak disorder, $\beta\ll 1$.
Even for the lowest LL, the SCBA density of states in two dimensions is parametrically correct.
\cite{PhysRevB.77.195430}
The exact shape can be calculated along the lines of Refs.~\onlinecite{Wegner, Brezin198424} In three dimensional systems for weak disorder,
the extra integration over the momentum $p_z$ further reduces the difference between the
exact and SCBA results in the limit $\varepsilon \to 0$.

\subsection{Energies close to the lowest Landau level}
\label{subsec:ImSigmaLLL}

We first consider the case of lowest energies, $|\varepsilon|\ll \Omega$ (i.e., $N_\varepsilon=0$) for weak disorder, $\beta\ll 1$.
In this case, the asymmetry with respect to the zeroth LL should be taken into account
and the imaginary parts of the two self-energies strongly differ from each other.
When the lowest Landau level is well separated from the others ($\text{Im}\Sigma_{1,2}\ll \Omega$),
the contribution of higher Landau levels to the sum over $n$ can be treated within the Born approximation,
while the contribution of $n=0$ should be calculated self-consistently. Within this procedure,
we immediately get $\text{Im}\Sigma_2=0$ and
\begin{equation}
\text{Im}\Sigma_1(\varepsilon) \simeq -A \int_{-\infty}^{\infty}dz \
\frac{\text{Im}\Sigma_1(\varepsilon)}{(\varepsilon-z)^2+[\text{Im}\Sigma_1(\varepsilon)]^2}=-A.
\label{ImSigmaLLL}
\end{equation}
This result coincides with the result of the non-self-consistent Born approximation, Sec. \ref{subsec:BA}.
The zeroth LL is separated from the first as long as the condition $A<\Omega$ is fulfilled.
The density of states for $\varepsilon\ll \Omega$ is finite and, to the leading order, is energy-independent.

Using Eq.~(\ref{ImSigmaLLL}), we find the leading non-vanishing term in $\text{Im}\Sigma_2$ for $\varepsilon\ll \Omega$:
\begin{eqnarray}
\text{Im}\Sigma_{2}(\varepsilon)& \simeq &
-\frac{\pi A^2}{2} \sum_{n=1} ^{N_\text{max}}\frac{\Omega^2 n}{(\Omega^2 n-\varepsilon^2)^{3/2}}
\sim -A \beta.
\label{Sigma2LLL}
\end{eqnarray}
Thus, in the limit of weak disorder, $\beta\ll 1$, we can neglect $\text{Im}\Sigma_{2}$ for $\varepsilon\ll \Omega$.
In fact, the condition is even softer: $\text{Im}\Sigma_{2}$ becomes of the order of $\text{Im}\Sigma_{1}$
only in the close vicinity of the first Landau level, $|\varepsilon-\Omega|\sim A$.

\subsection{Energies at high Landau levels}
\label{subsec:ImSigmaHLL}

We now consider high energies, $\varepsilon\gg \Omega$.
As we have seen in Sec. \ref{subsec:BA}, already within the Born
approximation the average (as well as minimal)
broadening of Landau levels increases with $\varepsilon$
parabolically, as in the zero-$H$ case: $\propto A (\varepsilon/\Omega)^2\sim \gamma \varepsilon^2/v^3$, see Fig. \ref{fig1}.
Therefore, the difference between $\text{Im}\Sigma_1$ and $\text{Im}\Sigma_2$
that comes from the contribution of $n=0$ can
be neglected for energies $\varepsilon$ away from the zeroth LL.
In what follows we set $\Sigma_{1}=\Sigma_{2}=\Sigma$ for $\varepsilon\gg \Omega$.

Introducing
\begin{equation}
\Gamma_{1,2}(\varepsilon)=-\text{Im}\Sigma_{1,2}(\varepsilon)
\label{Gamma12}
\end{equation}
and setting $\Gamma_1=\Gamma_2=\Gamma$,
we arrive at the self-consistent equation for $\Gamma(\varepsilon\gg\Omega):$
\begin{eqnarray}
\Gamma&=&\sum_{n=0} \Gamma^{(n)}(\varepsilon),
\label{GammaSCBA}
\\
\Gamma^{(n)}(\varepsilon)&=&\frac{A \Gamma}{\pi} \int_{-\infty}^\infty dz \frac{\varepsilon^2+\Omega^2 n + \Gamma^2}{(\varepsilon^2-\Omega^2 n - \Gamma^2-z^2)^2+4 \varepsilon^2 \Gamma^2}
\nonumber
\\
&=&A\ \text{Re} \frac{i\varepsilon + \Gamma}{\sqrt{W_n^2 -\varepsilon^2 + 2i \varepsilon \Gamma}},
\label{Gamman}
\end{eqnarray}
where we have introduced the partial contribution $\Gamma^{(n)}(\varepsilon)$ of the $n$th Landau level to the total
broadening $\Gamma(\varepsilon)$.
Note that each term in the r.h.s. of Eq.~(\ref{GammaSCBA})
contains the total broadening $\Gamma$ rather than the partial $\Gamma^{(n)}$.
Further, the position of the $n$th Landau level is shifted by disorder:
$\Omega^2 n$ appears only in combination
\begin{equation}
W_n^2=\Omega^2 n+\Gamma^2.
\label{Wn}
\end{equation}

\begin{figure}[t]
\centerline{\includegraphics[width=7cm]{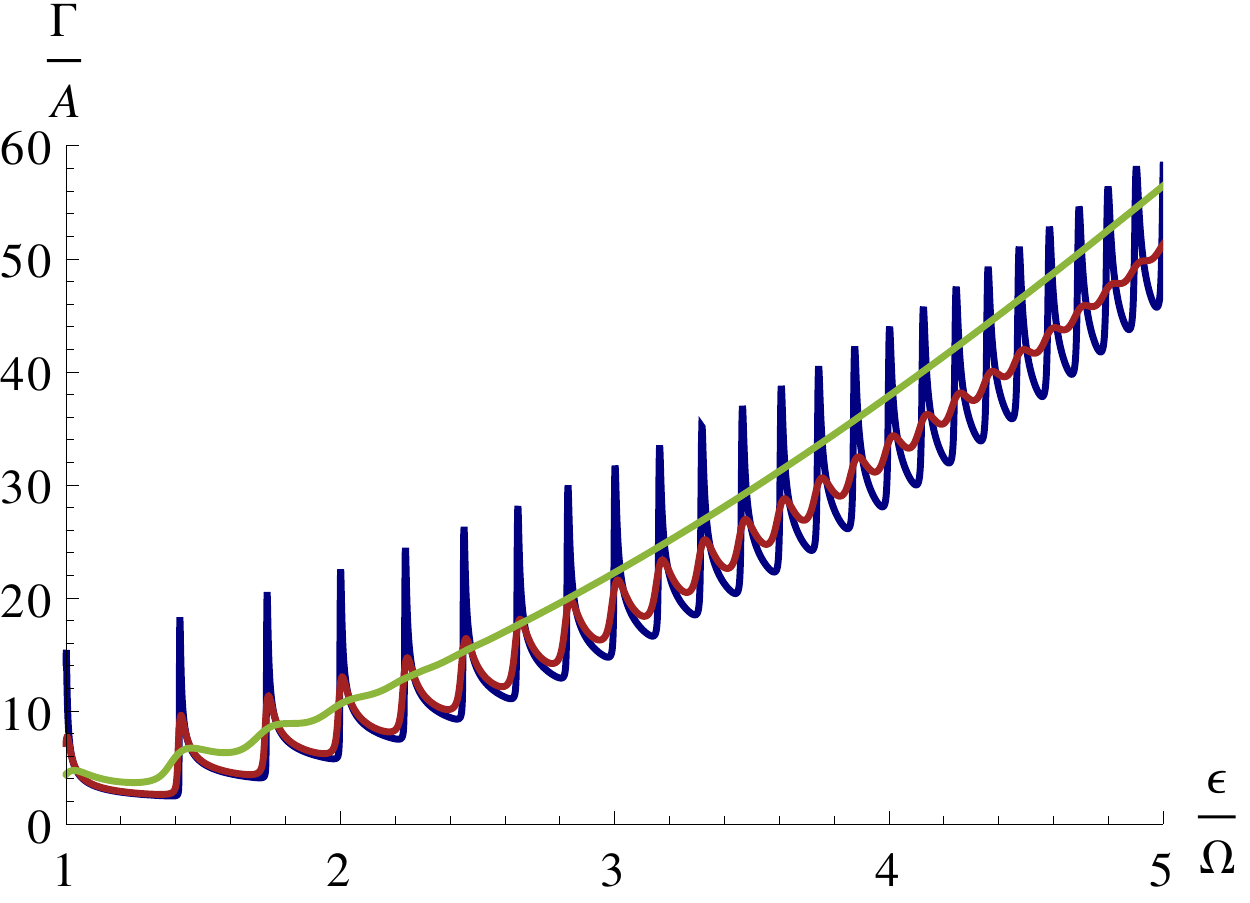}}
\caption{Disordered Weyl semimetal in the self-consistent Born approximation for $\varepsilon\gg \Omega$: $\Gamma(\varepsilon)$ in units of $A$ obtained by
numerical solution of Eq.~(\ref{GammaSCBA}).
Blue, red, and green curves correspond to $A/\Omega=10^{-4},\ 10^{-3},\ 10^{-2}$, respectively.
For all curves $N_{\text{max}}=100$.}
\label{fig2}
\end{figure}

In the case of weak disorder, for all energies $\Omega\ll \varepsilon \ll \Lambda$ we have $\varepsilon\gg \Gamma(\varepsilon)$,
so that Eq.~(\ref{Gamman}) can be written as
\begin{eqnarray}
\Gamma^{(n)}(\varepsilon)&\simeq &
A\varepsilon \frac{\sqrt{\varepsilon^2-W_n^2+\sqrt{(W_n^2-\varepsilon^2)^2+4 \varepsilon^2 \Gamma^2}}}{\sqrt{2}\  \sqrt{(W_n^2-\varepsilon^2)^2+4 \varepsilon^2 \Gamma^2} } .
\nonumber
\\
\label{GammanAPP}
\end{eqnarray}
For $\varepsilon\to \infty$ this yields $\Gamma^{(n)}(\varepsilon)\to A$ and for
$\varepsilon\ll W_n$ we get $\Gamma^{(n)}(\varepsilon)\to A \Gamma/W_n$.
Thus, when $\varepsilon$ crosses $W_n$, the $n$th Landau level gets an extra contribution $A$ to $\Gamma(\varepsilon)$.
The solution of the self-consistent equation (\ref{GammaSCBA}) is shown for $A/\Omega=10^{-4},\ 10^{-3},\ 10^{-2}$ in Fig. \ref{fig2}.

Let us now fix the Landau-level number $N\gg 1$ and consider the range of energies around $W_N$.
Assuming well separated Landau levels below $N$, we neglect $\Gamma$ in all terms with $n<N$:
\begin{equation}
\sum_{n=0}^{N-1} \Gamma^{(n)}(\varepsilon)\simeq A \sum_{n=0}^{N-1} \frac{\varepsilon}{\sqrt{\varepsilon^2-W_n^2}}
\simeq 2 A N.
\label{Gamman<N}
\end{equation}
The contribution of Landau levels with $n>N+1$ is dominated by $N_{\text{max}}$ and can be neglected for weak disorder:
\begin{equation}
\sum_{n=N+1} \Gamma^{(n)}(\varepsilon)\simeq A \sum_{n=N+1}^{N_\text{max}} \frac{W_n^2 \Gamma}{(W_n^2-\varepsilon^2)^{3/2}}
\sim \Gamma \beta \ll \Gamma.
\label{Gamman>N}
\end{equation}
Finally, the contribution of the $N$th Landau level (closest to the energy $\varepsilon$) can be further simplified for $|\varepsilon-W_N|\ll W_N$:
\begin{eqnarray}
\Gamma^{(N)}(\varepsilon)&\simeq &
\frac{A\ \sqrt{W_N}}{2}\ \frac{\sqrt{\varepsilon-W_N+\sqrt{(W_N-\varepsilon)^2+\Gamma^2}}}{\sqrt{(W_N-\varepsilon)^2+\Gamma^2} } .
\nonumber
\\
\label{GammaN}
\end{eqnarray}
In particular, exactly at $\varepsilon=W_N$ we find
\begin{equation}
\Gamma^{(N)}(\varepsilon=W_N)\simeq
\frac{A\varepsilon^{1/2}}{2\Gamma^{1/2}}.
\label{Gamma-maxN}
\end{equation}

Using Eqs.~(\ref{Gamman<N}) and (\ref{GammaN}), when $\varepsilon$ is close to $W_{N}$,
the self-consistency equation takes the form:
\begin{equation}
\Gamma(\varepsilon) \simeq  \frac{2 A \varepsilon^2}{\Omega^2} + \frac{A \sqrt{\varepsilon}}{2} \ \frac{\sqrt{\varepsilon-w_\varepsilon
+\sqrt{(w_\varepsilon-\varepsilon)^2+\Gamma^2(\varepsilon)}}}{ \sqrt{(w_\varepsilon-\varepsilon)^2+\Gamma^2(\varepsilon)} },
\label{SCEq}
\end{equation}
where
$w_\varepsilon\simeq \Omega\sqrt{N_\varepsilon}$,
so that the r.h.s. of Eq.~(\ref{SCEq}) explicitly depends on $\varepsilon$ only, as it should be.

Exactly at $\varepsilon=W_N$, the self-consistency equation reads:
\begin{equation}
\Gamma=\frac{2 A \varepsilon^2}{\Omega^2} + \frac{A\varepsilon^{1/2}}{2\Gamma^{1/2}}.
\label{SCNLL}
\end{equation}
We observe that for sufficiently small energies, the broadening is dominated by the self-consistent contribution
of the same Landau level, whereas for large energies, the broadening is given by the zero-$H$ result
stemming from lower Landau levels:
\begin{equation}
\Gamma(\varepsilon=W_N)\simeq
\begin{cases}
(A/2)^{2/3}\varepsilon^{1/3}, & \Omega\ll \varepsilon \ll \varepsilon_*,
\\
2 A (\varepsilon/\Omega)^2, & \varepsilon \gg \varepsilon_*,
\end{cases}
\label{Gamma-Center}
\end{equation}
where
\begin{equation}
\varepsilon_* \sim \Omega (\Omega/A)^{1/5} \propto \frac{H^{2/5}}{\gamma^{1/5}}.
\label{epsilon-star}
\end{equation}
Below $\varepsilon_*$ Landau levels are fully separated.
Each peak in $\Gamma(\varepsilon)$ is non-symmetric with respect to $W_N$, as inherited from the clean density of states.
The shape of the LL broadening is analyzed in detail in Appendix \ref{app:shape}.

For high energies, the behavior in zero magnetic field should be recovered.
Indeed, we can express the result for $\varepsilon>\varepsilon_*$ in terms of the energy as follows:
\begin{equation}
\Gamma(\varepsilon)=\frac{\gamma}{4\pi v^{3}}\varepsilon^{2}.
\end{equation}
We see that the magnetic field has dropped out from this result, as expected.
Thus the LL broadening is dominated by the $H=0$ result for $\varepsilon<\varepsilon_*$
In fact, taking into account the corrections to the broadening at $\varepsilon>\varepsilon_*$, we will
see in Sec.~\ref{sec:DOS} below that
the Landau level quantization of the density of states remains intact in a finite range of energies above $\epsilon_*$.
This should be contrasted with the 2D case, where a single scale separates regimes of strong and weak
Landau quantization. Finally, we note that magnetooscillations in Weyl semimetals were addressed
in Ref. \onlinecite{PhysRevB.87.245131} with phenomenological energy-independent broadening.
We find, however, that the energy dependence of $\Gamma$ is very rich.

\subsection{Strong Disorder}
\label{subsec:strong}

We now briefly discuss the regime of strong disorder, $\beta \gtrsim 1$.
As follows from the consideration of the weak-disorder case, see Eq.~(\ref{Sigma2LLL}),
for strong disorder the difference between the two self-energies, $\Sigma_1$ and $\Sigma_2$,
becomes inessential even at $\varepsilon=0$.
After the evaluation of the sum over $n$ in Eqs.~(\ref{Sigma1SCBA}) and (\ref{Sigma2SCBA}),
we find a qualitative change in the behavior of the imaginary part of self-energy (and thus of the density of states) at
$4A\sqrt{N_{max}}=\pi\Omega$.
This implies the existence of a critical disorder strength, $\gamma_{c,mag}=2\pi^2v^3/\Lambda$ separating
the two regimes.
 In the absence of magnetic field, the emergence of such a critical disorder strength $\gamma_c$ was reported in Refs.~\onlinecite{PhysRevB.89.054202}, \onlinecite{PhysRevLett.113.026602}, \onlinecite{PRL112-Herbut}, \onlinecite{PhysRevLett.114.257201} and \onlinecite{PhysRevLett.114.166601}. Remarkably, the critical disorder strength $\gamma_{c,mag}$ which we find for the case of a strong magnetic field turns out to be equal to the zero-field value $\gamma_c$.

The solution of the SCBA equation for strong disorder for $\varepsilon\gg \Omega$ is given by
\begin{equation}
\Gamma \simeq \frac{2\Omega\sqrt{N_\text{max}}}{\pi} -\frac{\Omega^2}{2A}
= 4\pi v^3\left(\frac{1}{\gamma_c}-\frac{1}{\gamma}\right),
\end{equation}
which is equal to the zero-$H$  result obtained in Ref.~\onlinecite{PhysRevB.89.054202}.
When $\gamma$ is substantially larger than $\gamma_c$ (i.e., $\gamma-\gamma_c \gtrsim \gamma_c$),
the broadening becomes of the order of the ultraviolet cut-off,
\begin{equation}
\Gamma \sim \Lambda,
\label{Gamma-strong}
\end{equation}
which ensures that all Landau levels overlap.
Further, at $\varepsilon\ll \Omega$, the solution of the SCBA equations yields for the LL broadening
\begin{equation}
\text{Im}\Sigma_{1} \simeq \text{Im}\Sigma_{2} \sim -\Omega\sqrt{N_{max}} \sim \Lambda.
\end{equation}
Thus, when the disorder is substantially stronger than the critical one, even the zeroth LL overlaps with the rest of the spectrum.

Within the SCBA, the real part of the self energy for $\beta=\gamma/\gamma_c\gg 1$ is found to be
\begin{equation}
\text{Re}\Sigma_{1} \simeq \text{Re}\Sigma_{2} \simeq \frac{\beta-2}{\beta-1}\varepsilon,
\label{Re-strong}
\end{equation}
yielding ${\tilde{\varepsilon}}\sim \varepsilon/\beta$.

In this paper, we do not discuss the critical regime near the transition from weak to strong disorder at $\beta\sim 1$.
At zero magnetic field, the criticality was addressed using the $\epsilon$-expansion within the renormalization group
approach in Refs.~\onlinecite{PhysRevLett.114.166601,PhysRevB.91.035133}.
The effect of magnetic field near the transition remains a very interesting question for future work.

\subsection{Density of States}\label{sec:DOS}

\begin{figure}
\begin{center}
\includegraphics[width=7cm]{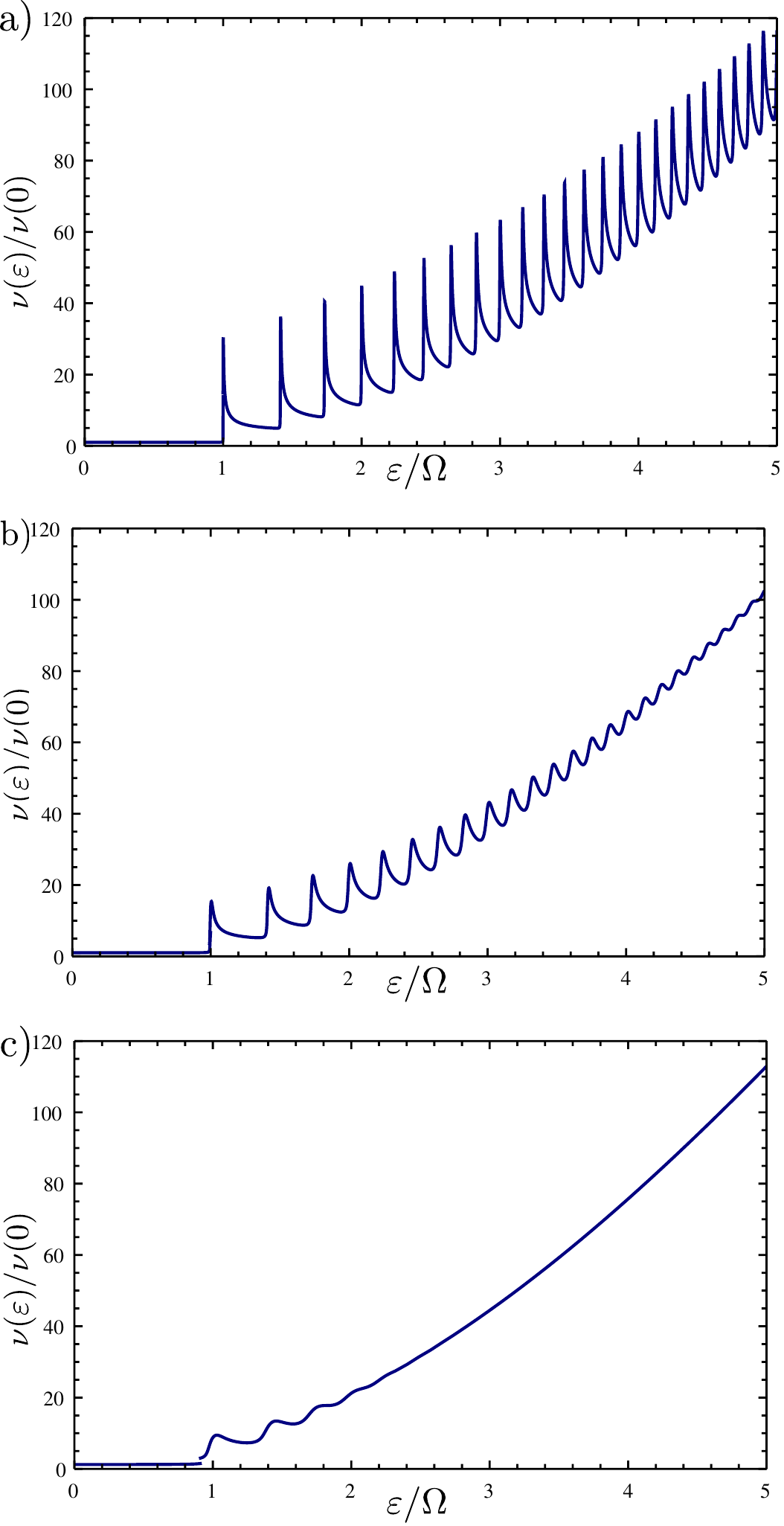}
\caption{Density of states, $\nu(\varepsilon)$, within self-consistent Born approximation in units of $\nu(0)\propto H$  [Eq.~\eqref{DOS-0}], as obtained from Eq.~(\ref{GammaSCBA}) for $\varepsilon\gtrsim \Omega$ and Eq.~(\ref{ImSigmaLLL}) for $\varepsilon<\Omega$.
The curves corresponds to (a) $A/\Omega=10^{-4}$, (b) $A/\Omega=10^{-3}$, and (c) $A/\Omega=10^{-2}$.
The value $N_{\text{max}}=100$ was used.}
\label{denslow}
\end{center}
\end{figure}

In Fig. \ref{denslow},
we plot the density of states obtained
by a numerical solution of the SCBA equation in the case of weak disorder.
The three figures illustrate the evolution of the density of states with the
increasing value of the parameter $\gamma\Omega/v^3$ (proportional to the disorder strength and to the square root of the magnetic field).

\begin{figure*}
\begin{center}
\includegraphics[width=16cm]{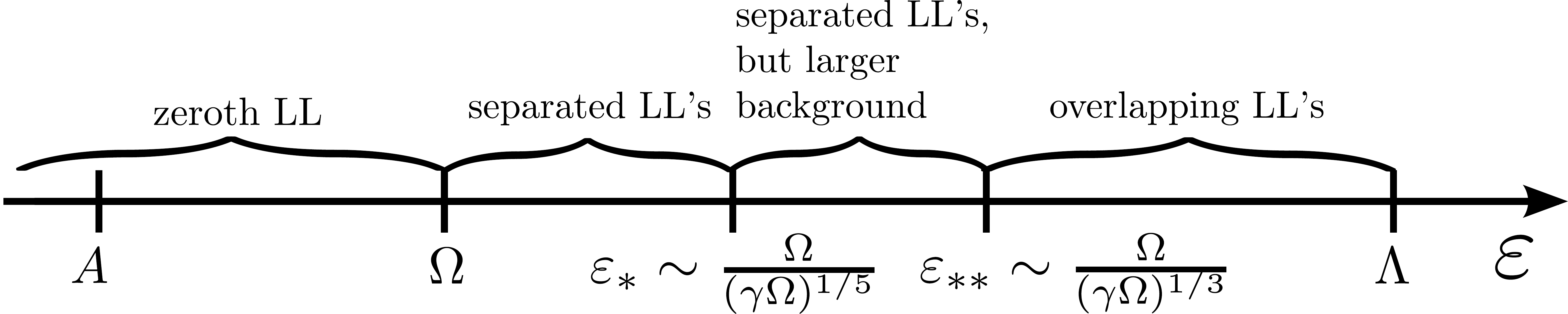}
\caption{Relevant energy scales and dominant contributions to the density of states of a disordered Weyl semimetal (for brevity, we have set $v=1$).}
\label{contributions}
\end{center}
\end{figure*}

Even in the clean case, Landau levels are broadened due to the integration over $p_z$, see Fig. \ref{density},
so that the divergent peaks at $\varepsilon_n(p_z)$ are located on top of the background density of states.
However, these peaks are well resolved for all energies.
Disorder leads to the suppression of the peaks and eventually Landau levels fully overlap at high energies.
Let us discuss the characteristic values of Landau level index
at which the behavior of the density of states changes qualitatively.

At $\varepsilon \agt \varepsilon_*$, the broadening of Landau levels is dominated by the background (zero-$H$) contribution.
From Eq.~(\ref{epsilon-star}), we see that the corresponding LL index $N_*=\varepsilon_*^2/\Omega^2$  decreases with increasing $H$:
\begin{equation}
N_*\sim (\Omega/A)^{2/5} \propto \frac{1}{\gamma^{2/5}H^{1/5}}.
\label{N-star}
\end{equation}

In order to find out whether LLs are resolved or not, we should check whether
the condition $\varepsilon_{n+1}(p_{z}=0)-\varepsilon_{n}(p_{z}=0)>\Gamma$ is fulfilled.
For energies $\Omega\ll \varepsilon \ll \Omega(\Omega/A)^{1/5}$, the width of high Landau levels
is smaller than the distance $\Omega^2/\varepsilon$ between them.
Therefore, at $\varepsilon\sim \varepsilon^*$ we have a situation when Landau levels are still resolved
on top of the background, but the height of the small peaks
$A^{1/2}\Omega/\varepsilon^{1/2}$, Eq.~(\ref{SCNLL}), is lower than the height of the background.
The latter then dominates the broadening.

The neighboring peaks fully overlap in the regime $\varepsilon>\varepsilon_{**}>\varepsilon_*$,
where the broadening is given by the zero-$H$ result:
\begin{equation}
A\frac{\varepsilon^{2}}{\Omega^2} \sim \frac{\Omega^2}{\varepsilon} \quad \Rightarrow \quad
\varepsilon\sim \varepsilon_{**}=\Omega\left(\frac{\Omega}{A}\right)^{1/3}.
\end{equation}
The corresponding LL index
\begin{equation}
N_{**}=\frac{\varepsilon_{**}^2}{\Omega^2} = \left(\frac{\Omega}{A}\right)^{2/3} \propto \frac{1}{\gamma^{2/3}H^{1/3}}
\end{equation}
decreases with increasing magnetic field, similarly to $N_{*}$.
Let us emphasize that, contrary to conventional expectations,
the number of separated LLs decreases with increasing magnetic field.
We thus see that a large number of low-lying Landau levels are well resolved for weak magnetic field.

When the magnetic field increases for a fixed disorder strength, the Landau level index $N_{**}$ associated with the starting point of overlapping, becomes smaller. This behavior is very unusual, as it is opposite to that in the case of conventional semiconductors. The energy $\varepsilon_{**}$ where the Landau levels start to overlap
increases with $H$ as $H^{1/3}$.
With regard to the behavior with disorder strength, the obtained results qualitatively
conform with intuitive expectations.
Specifically, with increasing disorder, the number of separated Landau levels decreases and the corresponding energy range shrinks.

It is important to stress that, in the presence of magnetic field, the density of states at the Dirac  point (zero energy) is finite even in the weak-disorder regime. Specifically, the value of the density of states at $\varepsilon=0$ is linear in magnetic field:
\begin{equation}
\nu(0)=\frac{A}{\pi \gamma}=\frac{\Omega^2}{8\pi^2 v^3} \propto H.
\label{DOS-0}
\end{equation}
It is worth noting that a finite value of the density of states at the degeneracy point will lead to a finite conductivity independently
of the order of limits $\omega\to 0$ and $T\to 0$.
From this point of view, a finite magnetic field has the same effect as a strong disorder.

All the features of the density of states that we have found analytically (see Fig.~\ref{contributions}) are perfectly observed in Figs. \ref{denslow} a,b,c. First, one sees that for weak disorder and weak magnetic field many LLs are separated and that the number of separated Landau levels decreases with increasing magnetic field or with increasing disorder. Second, there is an intermediate range of energies where the density of states is dominated by the background
value but Landau levels are well resolved. Third, one observes that the background density of states is equal to that in the absence of magnetic field (quadratic in energy).
Finally, the magnetic field creates a finite density of states at the degeneracy point which depends on the magnetic field.

\section{Conductivity at charge neutrality}
\label{sec:conductivity}

In this Section, we calculate the conductivity of a disordered Weyl semimetal in the presence of a quantizing transversal magnetic field.
Here we restrict ourselves to the case of weak disorder and to zero chemical potential, $\mu=0$.
We use the Kubo formula for the real part of the longitudinal conductivity,
\begin{align}\label{Kubo}
&\sigma_{xx}(\omega,T)=\int\frac{d\varepsilon}{2\pi}\frac{f_{T}(\varepsilon)}{\omega}\int\frac{d^{3}\textbf{p}}{(2\pi)^{3}}\nonumber\\
&\quad \times \text{Tr}\left\{\left[\hat{G}^{R}(\varepsilon,\textbf{p})-\hat{G}^{A}(\varepsilon,\textbf{p})\right]\hat{j}_{x}^\text{tr}
\hat{G}^{A}(\varepsilon-\omega,\textbf{p})\hat{j}_{x}\right.\nonumber\\
&\quad \left.+\hat{G}^{R}(\varepsilon+\omega,\textbf{p})\hat{j}_{x}^\text{tr}\left[\hat{G}^{R}(\varepsilon,\textbf{p})
-\hat{G}^{A}(\varepsilon,\textbf{p})\right]\hat{j}_{x}\right\}.
\end{align}
Here $\hat{j}_x=ev\sigma_x$ is the bare current operator and $\hat{j}_x^\text{tr}= V^\text{tr} \hat{j}_x$ is the current vertex dressed by disorder,
see Appendix~\ref{app:vertcorr}.
The effect of disorder manifests itself in the replacement of bare Green's functions by
impurity-averaged matrix Green's functions \eqref{green} and in the appearance of the current vertex corrections $V^\text{tr}$.
As discussed in Ref.~\onlinecite{PhysRevB.89.014205}, the calculation of the conductivity in Weyl semimetals requires
taking into account vertex corrections even for point-like disorder (similarly to graphene).
In the absence of magnetic field, the inclusion of vertex corrections away from the Weyl point leads to the difference
between the transport and quantum (coming from the single-particle self-energy) scattering times:~\cite{PhysRevB.89.014205} $\tau^\text{tr}=(3/2)\tau^q$.
In what follows, we first evaluate the conductivity without vertex corrections and then include the vertex corrections at the end of the calculation.

Starting from Eq.~\eqref{Kubo}, setting $V^\text{tr}=1$, evaluating the trace,
and taking into account the orthogonality of wave functions of different LLs,
we find that the Kubo formula for the conductivity without vertex corrections takes the following form in the LL representation:
\begin{align}
\sigma_{xx}^{(0)}\left(T\right)&=\frac{e^{2}v^{2}}{T}\int\frac{d\varepsilon}{2\pi}\frac{1}{\cosh^{2}\left(\frac{\varepsilon-\mu}{2T}\right)}
\  \sum_{n} \frac{eH}{2\pi c}
\notag
\\
&\times
\int\frac{dp_{z}}{2\pi}\ \text{Im} G^{R}_{11}(\varepsilon,n,p_{z})\ \text{Im}G^{R}_{22}(\varepsilon,n,p_{z}).
\label{transversal}
\end{align}
Since the self-energies for the zeroth Landau level differ from those for higher Landau levels,
we have to distinguish between the zeroth Landau level and the others. This is also true for the case of vertex corrections.
For low temperature, $T<\Omega$, the conductivity is dominated by the contribution of the zeroth LL.
For higher temperatures, excitations to higher LLs are possible and therefore,
the conductivity is determined by the contributions of the zeroth LL, separated and overlapping LLs.

\subsection{Low temperatures $T\ll \Omega$: Zeroth Landau Level}

We consider first the situation when the contribution of the zeroth LL is dominant. This is the case under the following two assumptions:
(i) the zeroth LL is separated from the first one, which is the case if the condition $A<\Omega$ is fulfilled;
(ii) the temperature satisfies  $T<\Omega$, so that excitations to higher LLs are suppressed exponentially.
In this case, the integral over energy $\varepsilon$
is dominated by the contribution of the zeroth Landau level.
We note that for $\Omega\ll \Lambda$ and weak disorder $\gamma<\gamma_c$ there is no room for the regime $A>\Omega$, since this would imply
$\gamma \Omega/v^3 > 1$ whereas $\gamma \Lambda/v^3<1$.
Furthermore, the current vertex correction for energies close to the Weyl node turn out to be small, $V^{\text{tr}}(\varepsilon\ll \Omega)\sim A/\Omega\ll 1$, see Appendix~\ref{app:vertcorr}. Therefore, in the regime of the dominant zeroth LL contribution, we will ignore the difference between the quantum scattering time and transport scattering time.

Using $\text{Im}\Sigma_1\simeq A$ and $\text{Im}\Sigma_2\simeq 0$ and disregarding the real part of self-energies,
we get
\begin{eqnarray}
G^{R}_{11}(\varepsilon,n,p_{z})&\simeq& \frac{\varepsilon+v p_z}{(\varepsilon+i A-v p_z)(\varepsilon+v p_z)-\Omega^2 n},
\label{G11e0}
\\
G^{R}_{22}(\varepsilon,n,p_{z})&\simeq& \frac{\varepsilon+iA-v p_z}{(\varepsilon+i A-v p_z)(\varepsilon+v p_z)-\Omega^2 (n+1)}.
\nonumber
\\
\label{G22e0}
\end{eqnarray}
Substituting Eqs.~(\ref{G11e0}) and (\ref{G22e0}) into Eq.~(\ref{transversal}) and setting $\varepsilon=0$ in Green's functions,
we arrive at ($z=v p_z$):
\begin{align}
\sigma_{xx}\left(T\right)&\simeq \frac{e^{2} A^2 \Omega^4}{2\pi^2 v }\ \sum_{n=0}
\int\frac{dz}{2\pi}  \frac{z^2}{[(z^2+\Omega^2 n)^2+A^2 z^2]} \nonumber
\\
&\quad \times
\frac{(n+1)}{\{[z^2+\Omega^2 (n+1)]^2+A^2 z^2\}}.
\label{cond-LLL}
\end{align}
Neglecting $A\ll \Omega$ (the condition of separation
of the lowest Landau level) in the denominators for
higher Landau levels,
we see that the sum over $n>0$ converges and gives the contribution $\sim e^2A^2/(\Omega v)$ to the
conductivity, whereas the $n=0$ term yields $e^2A/v$.
Thus, for $A\ll \Omega$ the total conductivity at $\mu=0$ and $T\ll \Omega$ is dominated
by the contribution of the zeroth Landau level and is
given by:
\begin{align}
\sigma_{xx}(T\ll \Omega \ll v^3/\gamma)&\simeq
\frac{e^2}{(2\pi)^2}\frac{A}{v} = \frac{e^2}{16 \pi^3}\frac{\gamma \Omega^2}{v^4} \propto \gamma H.
\label{cond-0-result}
\end{align}
The resulting conductivity Eq.~\eqref{cond-0-result} is proportional both to the disorder strength and to the magnetic field.

\subsection{High temperatures, $T\gg \Omega$}
\label{cond-highT}

For higher temperatures, $T\gg \Omega$, energies $\varepsilon\gg \Omega$ are involved in the thermal averaging,
so that we need to evaluate the contribution of high Landau levels to the conductivity.
For $\varepsilon\gg \Omega$ we neglect the difference between the self-energies: $\Sigma_1=\Sigma_2$.
As before, we include the real part of self-energies into the shifted energy $\tilde{\varepsilon}$
and drop the tilde everywhere. The imaginary part of the self-energy is written through the Landau-level broadening:
$\text{Im}\Sigma(\varepsilon)=-i \Gamma(\varepsilon).$

The Green functions for $\varepsilon\gg \Omega$ take the form:
\begin{eqnarray}
G^{R}_{11}(\varepsilon,n,p_{z})&\simeq& \frac{\varepsilon+v p_z+i\Gamma}{(\varepsilon+i \Gamma)^2-v^2 p_z^2-\Omega^2 n},
\label{G11n}
\\
G^{R}_{22}(\varepsilon,n,p_{z})&\simeq& \frac{\varepsilon+v p_z+i\Gamma}{(\varepsilon+i \Gamma)^2-v^2 p_z^2-\Omega^2 (n+1)},
\label{G22n}
\end{eqnarray}
yielding with $z=p_z v$
\begin{eqnarray}
\text{Im}G^{R}_{11}&\simeq& -\Gamma \frac{\varepsilon^2+z^2+\Omega^2 n+\Gamma^2+2 \varepsilon z}{(\varepsilon^2-z^2-\Omega^2 n-\Gamma^2)^2+4\varepsilon^2\Gamma^2},
\label{ImG11n}
\\
\text{Im}G^{R}_{22}&\simeq&
-\Gamma \frac{\varepsilon^2+z^2+\Omega^2 (n+1)+\Gamma^2+2 \varepsilon z}{(\varepsilon^2-z^2-\Omega^2 (n+1)-\Gamma^2)^2+4\varepsilon^2\Gamma^2}.
\nonumber
\\
\label{ImG22n}
\end{eqnarray}
Substituting these in Eq.~(\ref{transversal}), we arrive at
\begin{eqnarray}
\sigma_{xx}^{(0)}&=&\frac{e^2\Omega^2}{2\pi^2 v} \int_{-\infty}^\infty \frac{d\varepsilon}{4T \cosh^{2}\left(\frac{\varepsilon}{2T}\right)}
\sum_{n=0} Q_n(\varepsilon),
\label{sigmaQ}
\\
Q_n(\varepsilon)&=&\int_{-\infty}^\infty \frac{dz}{2\pi}\ \text{Im}G^{R}_{11}(\varepsilon,n,p_{z}) \text{Im}G^{R}_{22}(\varepsilon,n,p_{z}).
\nonumber
\\
\label{Qne}
\end{eqnarray}
The evaluation of the integral in Eq.~\eqref{Qne} then yields
\begin{eqnarray}
Q_n(\varepsilon)
&=& \frac{\Gamma}{2} \text{Re}
\left\{
\left[\frac{1}{\sqrt{\varepsilon^2-\Omega^2 (n+1)-\Gamma^2-2i\varepsilon \Gamma}}
\right.
\right.
\label{QnFull}
\\
&+&\left.\left.
\frac{1}{\sqrt{\varepsilon^2-\Omega^2 n-\Gamma^2+2i\varepsilon \Gamma}}\right]
\frac{\varepsilon(2n+1)+i\Gamma}{\Omega^2+  4 i \varepsilon \Gamma}
\right\}.
\nonumber
\end{eqnarray}

Let us now include the vertex corrections. The total conductivity is then given by Eq.~(\ref{sigmaQ}) with the replacement
$Q_n \to Q_n^\text{tr}$, where $Q_n^\text{tr}$ includes the dressing of the current operator by disorder (``transportization'').
The vertex correction $V^\text{tr}(\varepsilon \gg \Omega)$ is calculated in Appendix~\ref{app:vertcorr}:
\begin{equation}
V^\text{tr}\simeq \frac{\Omega^2+4i\varepsilon \Gamma}{\Omega^2+\frac{8}{3}i\varepsilon \Gamma}.
\label{Vtr}
\end{equation}
The inclusion of vertex corrections replaces $4i\varepsilon\Gamma$ with $(8/3)i\varepsilon\Gamma$ in the denominator
of Eq.~(\ref{QnFull}). In zero magnetic field this yields $\sigma_{xx}=(3/2)\sigma_{xx}^{(0)}$,
in agreement with Ref.~\onlinecite{PhysRevB.89.014205}. Below we will see that the effect of vertex corrections in magnetic field
is captured by the replacement of $\tau^q$ with $\tau^\text{tr}=(3/2)\tau^q$ in the Drude-like formula for the magnetoconductivity.

A detailed evaluation of $Q_n$ for $T\gg \Omega$ is given in Appendix \ref{app:highT}.
For $\varepsilon>\varepsilon^*$, when the LL broadening is dominated by the background, $\Gamma=2A \varepsilon^2/\Omega^2$,
we have
\begin{equation}
\sum_{n=0} Q_n^\text{tr} \simeq \frac{4\Gamma \varepsilon^4}{\Omega^2[(4 \varepsilon \Gamma)^2+9\Omega^4/4]}.
\label{sumQn-result}
\end{equation}
As a result, the contribution of this energy region to the conductivity reads:
\begin{eqnarray}
\sigma_{xx}& \simeq &
\frac{e^2}{\pi^2}\ \frac{A \Omega^2}{ v T }
\int \frac{d\varepsilon}{ \cosh^{2}\left(\frac{\varepsilon}{2T}\right)}
\frac{\varepsilon^6}{(8A\varepsilon^3)^2+9\Omega^8/4}.
\label{sigmaxx-class}
\end{eqnarray}
This expression can be cast in the form of a conventional Drude-like formula for the magnetoconductivity
(for recent review see Ref.~\onlinecite{RMP})
with the $\varepsilon$-dependent transport scattering
time $\tau^\text{tr}(\varepsilon)$ and effective cyclotron frequency $\omega_c(\varepsilon)$,
\begin{equation}
\sigma_{xx}^D=\frac{e^2v^2}{6\pi}
\int \frac{d\varepsilon}{4T \cosh^{2}\left(\frac{\varepsilon}{2T}\right)}\
\frac{\nu(\varepsilon)\tau^\text{tr}(\varepsilon)}{1+\omega_c^2(\varepsilon)[\tau^\text{tr}(\varepsilon)]^2}.
\label{Drude}
\end{equation}
Indeed, using the self-consistency relation between the densities of states and scattering times in magnetic field
\begin{equation}
\nu(\varepsilon)\tau^\text{tr}(\varepsilon)=\frac{3}{4\pi \gamma},
\label{SCBA-rel}
\end{equation}
the semiclassical expression for the cyclotron frequency in the linear spectrum
\begin{equation}
\omega_c(\varepsilon)=\frac{v^2}{l_H^2 \varepsilon}=\frac{\Omega^2}{2\varepsilon},
\label{cyclotron}
\end{equation}
and expressing the quantum scattering rate through the broadening
\begin{equation}
\Gamma(\varepsilon)=2A \frac{\varepsilon^2}{\Omega^2}=\frac{1}{2\tau^q(\varepsilon)}=\frac{3}{4\tau^\text{tr}(\varepsilon)},
\end{equation}
we obtain Eq.~(\ref{sigmaxx-class}) from Eq.~(\ref{Drude}).

Furthermore, as demonstrated in Appendix \ref{app:highT}, in the energy range $\Omega<\varepsilon<\varepsilon_*$,
where Landau levels are separated, the conductivity is still dominated by the low-lying LLs
with $n<N_\varepsilon$, leading to Eq.~\eqref{sigmaxx-class} also in this regime.
It thus turns out that
the semiclassical Drude formula (\ref{Drude}) is valid for all temperatures $T\gg \Omega$.
A similar result was obtained for graphene at the charge neutrality point in
Refs. \onlinecite{PhysRevB.87.165432} and \onlinecite{PhysRevB.87.195432}.

Let us now analyze the magnetic-field dependence of the conductivity
for $T>\Omega$.
The denominator of the integrand in Eq.~(\ref{sigmaxx-class})
is dominated by $\Omega^2$ for $\varepsilon<\varepsilon_{**}=\Omega(\Omega/A)^{1/3}$.
Therefore, for $\Omega<T<\varepsilon_{**}$ we obtain
\begin{eqnarray}
\sigma_{xx}& \simeq & \frac{4e^2}{9\pi^2}\ \frac{A \Omega^2}{ v T }
\int \frac{d\varepsilon}{ \cosh^{2}\left(\frac{\varepsilon}{2T}\right)}
\frac{\varepsilon^6}{\Omega^8}
= \frac{62 \pi^3}{189} \frac{e^2 \gamma T^6}{v^4 \Omega^4} \propto \frac{1}{H^2}.
\nonumber
\\
\label{sigmaxx-*-**}
\end{eqnarray}
For higher temperatures, $T>\varepsilon_{**}$, we neglect $\Omega^8$
in the denominator of the integrand in Eq.~(\ref{sigmaxx-class}), which yields
 \begin{eqnarray}
\sigma_{xx}& \simeq & \frac{e^2}{\pi^2}\ \frac{A \Omega^2}{ v T }
\int \frac{d\varepsilon}{ \cosh^{2}\left(\frac{\varepsilon}{2T}\right)}
\frac{\varepsilon^6}{(8 A \varepsilon^3)^2}
=  \frac{e^2 v^2}{2\pi \gamma}.
\nonumber
\\
\label{sigmaxx**}
\end{eqnarray}
This is just the conductivity in the absence of magnetic field.
The $H$ dependent correction to this result is non-analytic in $H$:
\begin{eqnarray}
\delta\sigma_{xx}\simeq
- \frac{e^2 v^3}{4} \left(\frac{\pi}{18}\right)^{1/3}\frac{\Omega^{2/3}}{\gamma^{4/3}T} \propto - \frac{H^{1/3}}{\gamma^{4/3} T}.
\nonumber
\\
\label{correction}
\end{eqnarray}
This correction will determine the low-field magnetoresistance.

\begin{figure}
\begin{center}
\includegraphics[width=8.5cm]{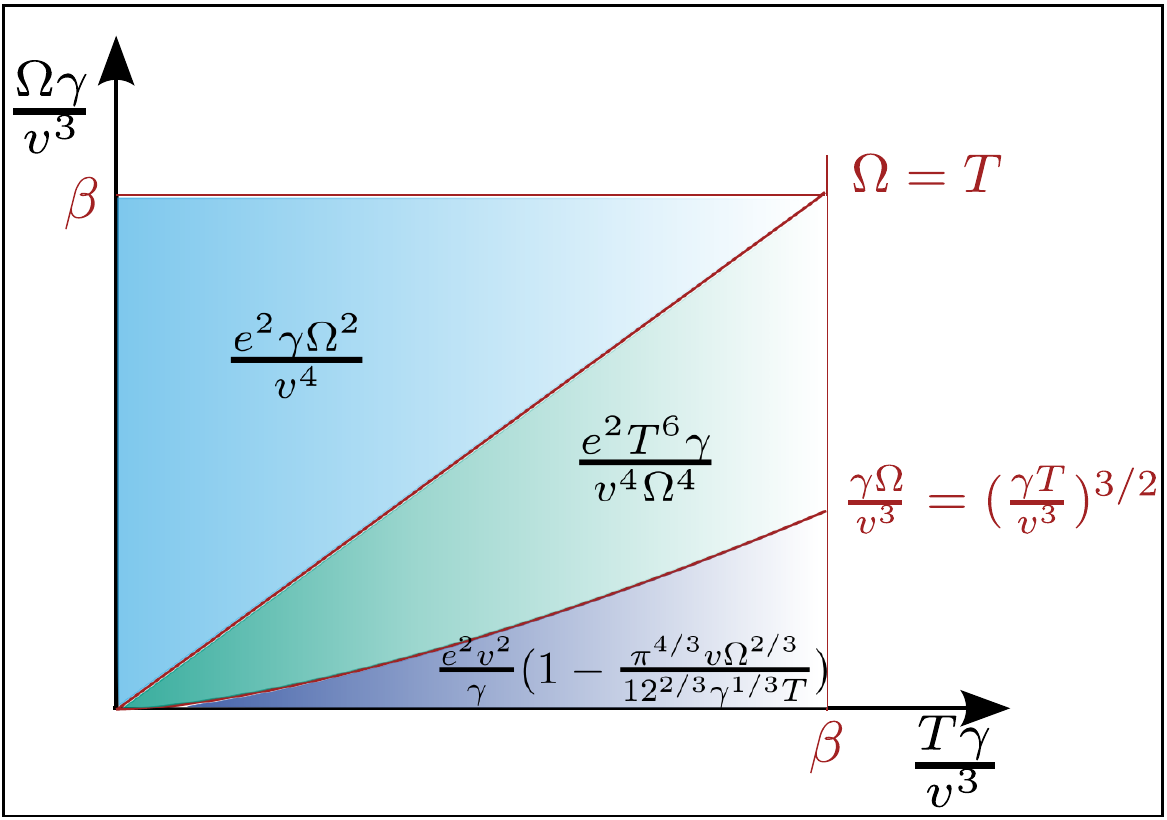}
\caption{
Behavior of the conductivity $\sigma_{xx}$ in a Weyl semimetal with weak white-noise disorder in dependence of temperature and magnetic field. Scaling of dominant contribution to the conductivity in each of the parameter regimes is shown.
Equations describing borderlines between the regimes are also indicated.
}
\label{phasediagramm}
\end{center}
\end{figure}

The summary of all regimes in the temperature-magnetic field plane is visualized
in Fig.~\ref{phasediagramm}. For each of the regimes, the scaling of the corresponding dominant contribution(s)
to the conductivity is shown.
The conductivity is dominated by the Drude formula (\ref{Drude}) down to the lowest Landau level,
$T\sim \Omega$:
\begin{equation}
\sigma_{xx}\sim
\left\{
  \begin{array}{ll}
    \dfrac{e^2 \gamma \Omega^2}{v^4} \propto H, &\quad  T \ll \Omega , \\[0.3cm]
    \dfrac{e^2 \gamma T^6}{v^4 \Omega^4}\propto \dfrac{1}{H^2}, &\quad
\Omega \ll T \ll \varepsilon_{**}=\dfrac{v \Omega^{2/3}}{\gamma^{1/3}} , \\[0.3cm]
    \dfrac{e^2 v^2}{\gamma}, &\quad  T\gg \varepsilon_{**}.
  \end{array}
\right.
\label{short-range-sigma}
\end{equation}
In the last regime $T\gg \varepsilon_{**}$, the correction to the conductivity
is given by Eq.~(\ref{correction}).

As is seen in Fig.~\ref{phasediagramm}, the limits $H\rightarrow0$ and $T\rightarrow0$
are not interchangeable for weak disorder. Specifically, setting first $H\rightarrow0$ and then $T\rightarrow0$ yields
\begin{equation}
\lim\limits_{T\to 0}\lim\limits_{H\to 0}\sigma_{xx}(H,T)=\frac{e^2 v^2}{2\pi \gamma}.
\end{equation}
On the other hand, if we first perform the limit $T\rightarrow0$ and then $H\rightarrow0$, we obtain
\begin{equation}
\lim\limits_{H\to 0}\lim\limits_{T\to 0}\sigma_{xx}(H,T)=0.
\end{equation}
This demonstrates a peculiarity of the transport properties at the Weyl point.
As  has been mentioned in the Introduction, a similar behavior was also found in the absence of magnetic field,
Ref.~\onlinecite{PhysRevLett.108.046602}. Specifically, at $H=0$ it was crucial to distinguish between the order of the limits $T\rightarrow0$ and $\omega\rightarrow0$ for weak disorder, while in the strong disorder regime, the order of the limits was interchangeable
(see Refs.~\onlinecite{PhysRevB.84.235126} and \onlinecite{PhysRevLett.108.046602}).
In the presence of magnetic field, the limits $\omega\rightarrow0$ and $T\rightarrow0$ become interchangeable
also for weak disorder because of a finite density of states at the Weyl node.
The reason for the elimination of the non-interchangeability of two limits by magnetic field (for any disorder strength)
or by strong disorder is the generation of a finite density of states at the Dirac point ($\varepsilon=0$).

\subsection{Conductivity for strong disorder}

We now briefly discuss the magnetoconductivity in the case of strong disorder, $\beta=\gamma/\gamma_c\sim \gamma \Lambda/v^3\gg 1$.
As discussed in Sec. \ref{subsec:strong}, in this case the broadening of all Landau levels is of the order of the ultraviolet cut-off:
all Landau levels overlap. To calculate the conductivity, we use the semiclassical Drude formula (\ref{Drude}) complemented by
Eqs.~(\ref{Gamma-strong}), (\ref{Re-strong}), (\ref{SCBA-rel}), and (\ref{cyclotron}).
The parameter governing the magnetic-field dependence of the conductivity now reads
\begin{equation}
\omega_c(\tilde\varepsilon)\tau(\tilde\varepsilon)\sim \frac{\Omega^2}{\Gamma^2}\sim \frac{\Omega^2}{\Lambda^2}.
\end{equation}
Here we have replaced the energy $\tilde\varepsilon$ by the level broadening $\Gamma\sim\Lambda$.
The main contribution to the conductivity is independent of magnetic field,
whereas a weak magnetic field yields a quadratic-in-$H$ correction (here we do not write numerical prefactor in the $H$-dependent correction)
\begin{eqnarray}
\sigma_{xx}&\sim&\frac{e^2 v^2}{\gamma_c}
\int \frac{d\varepsilon}{T} \frac{1}{1+\Omega^4/\Lambda^4}
\sim \frac{e^2v^2}{\gamma_c}\left(1-\frac{\Omega^4}{\Lambda^4}\right),
\nonumber
\\
\delta\sigma_{xx} &\sim& -\frac{e^2\Omega^4}{v \Lambda^3}\propto - H^2.
\label{Drude-strong-2}
\end{eqnarray}
For strong disorder, the $H$-dependent conductivity correction vanishes quadratically. This should be
 contrasted with the conductivity for weak disorder which is non-analytic (proportional to $H^{1/3}$).

\section{Charged Impurities}
\label{sec:CoulombImp}

\subsection{Screening}

In the preceding part of the paper we considered a model of white-noise disorder.
We are now going to generalize the obtained results onto the more realistic case of screened Coulomb impurities.
The potential of such an impurity is given by
\begin{equation}
U(\textbf{k})=\frac{4\pi e^{2}}{\epsilon_{\infty}(k^{2}+\kappa^{2})},
\end{equation}
where $\epsilon_{\infty}$ is the background dielectric constant. The parameter $\kappa$ is the inverse Debye screening radius,
\begin{equation}
\kappa^{2}=\frac{4\pi e^2}{\epsilon_{\infty}}\frac{\partial n}{\partial\mu}
=\frac{e^{2}}{\pi\varepsilon_{\infty}v^3}
\left\{
  \begin{array}{ll}
    \Omega^2, &\quad \Omega\gg T, \\[0.2cm]
    \pi^2 T^2/3, &\quad \Omega\ll T.
  \end{array}
\right.
\end{equation}
Here $\partial n/\partial \mu$ is the fermion compressibility
and we have neglected the effect of disorder on the thermodynamic density of states.
In the limit $T,H \to 0$ one should include the impurity contribution in a self-consistent way,
as we will discuss below.

In the following, we shall assume that the ``fine-structure'' constant is not small,
\begin{equation}
e^2/v\agt 1.
\label{fine}
\end{equation}
In a realistic situation it is of order of unity, so that
$\kappa$ is of the order of characteristic values $k_{\text{typical}}\sim \text{max}(\Omega, T)/v$
of the wave vector $k$.
Condition (\ref{fine}) allows one to describe the screened Coulomb disorder by an effective
pointlike correlator
 \begin{equation}\label{impcorr}
\left\langle U(\textbf{r})U(\textbf{r}')\right\rangle\simeq\gamma(H,T)\delta(\textbf{r}-\textbf{r}')
\end{equation}
in order to find the parametric dependence of the conductivity (without numerical prefactors).

The correlator (\ref{impcorr}) corresponds to a white-noise disorder whose strength
depends on magnetic field and temperature,
 \begin{equation}\label{impcornew}
\gamma(H,T)=N_{\text{imp}}\left(\frac{\partial n}{\partial \mu}\right)^{-2}
\sim N_{\text{imp}}v^6
\left\{
  \begin{array}{ll}
    \Omega^{-4}, &\quad \Omega\gg T, \\[0.2cm]
    T^{-4}, &\quad \Omega\ll T,
  \end{array}
\right.
\end{equation}
where $N_{\text{imp}}$ is the density of impurities.
In the limit $T,H\to 0$, Eq.~(\ref{impcornew}) yields a divergent disorder strength
which implies the necessity of a
self-consistent treatment of the impurity screening.
Specifically, when at
\begin{equation}
\text{max}(\Omega,T) \sim \varepsilon_\text{imp}=N_{\text{imp}}^{1/3}v
\end{equation}
the quasiparticle broadening $\gamma(H,T)\text{max}(T^2,\Omega^2)/v^3$  becomes of the
order of $\text{max}(\Omega,T)$, the screening will be determined by the
impurity-induced density of states, yielding
\begin{equation}
\gamma(H,T)\sim \gamma_0=N_{\text{imp}}^{-1/3}v^2 \ll \gamma_c.
\end{equation}
The weak-disorder approach is applicable
under the condition $\text{max}(\Omega,T) \agt \varepsilon_\text{imp}$.

Below we employ the results of the previous sections
to the Coulomb case by replacing $\gamma$ with $\gamma(H,T)$.
Using Eq.~(\ref{impcornew}), we express the condition for separation of the zeroth LL, $A<\Omega$, as
\begin{equation}
\Omega > v^3/\gamma_0 = \varepsilon_\text{imp}.
\end{equation}
We see that for screened Coulomb impurities the increase of magnetic field favors the Landau quantization,
as opposed to the case of white-noise disorder.
In particular, in the limit $H\to \infty$, the zeroth LL is always separated.
This demonstrates a crucial role played by the $H$ dependence of the screening.

\subsection{Conductivity}

In order to find the conductivity, we should substitute $\gamma(H,T)\sim \gamma_0^{-3}[\text{max}(\Omega,T)]^{-4}$ for $\gamma$
in the results for the conductivity of  a system with white-noise disorder. Since we do not keep numerical factors, we
also disregard the vertex corrections (that only modify these factors, see above).
We begin by considering the low-temperature regime, $T\ll \Omega$, when the screening is controlled by the magnetic field.
If $\Omega < \varepsilon_\text{imp}$, all LLs overlap, and the conductivity is essentially equal to that at zero magnetic field, see below.
In the opposite situation, $\Omega>\varepsilon_\text{imp}$, only the zeroth LL contribute. With $A\sim N_\text{imp}v^3/\Omega^2$, this yields
\begin{equation}\label{lowtcoul}
\sigma_{xx}\sim \frac{e^2 A}{v} \sim \frac{e^2 \varepsilon_\text{imp}^3}{v \Omega^2}\sim  \frac{e^2 N_\text{imp} v^2}{\Omega^2} \propto \dfrac{1}{H}.
\end{equation}
This result agrees with the result obtained by Abrikosov in Ref.~\onlinecite{PhysRevB.58.2788}.
We observe that the dependence of conductivity on the magnetic field for Coulomb impurities differs strongly from that for white-noise impurities.

\begin{figure*}
\begin{center}
\includegraphics[width=16cm]{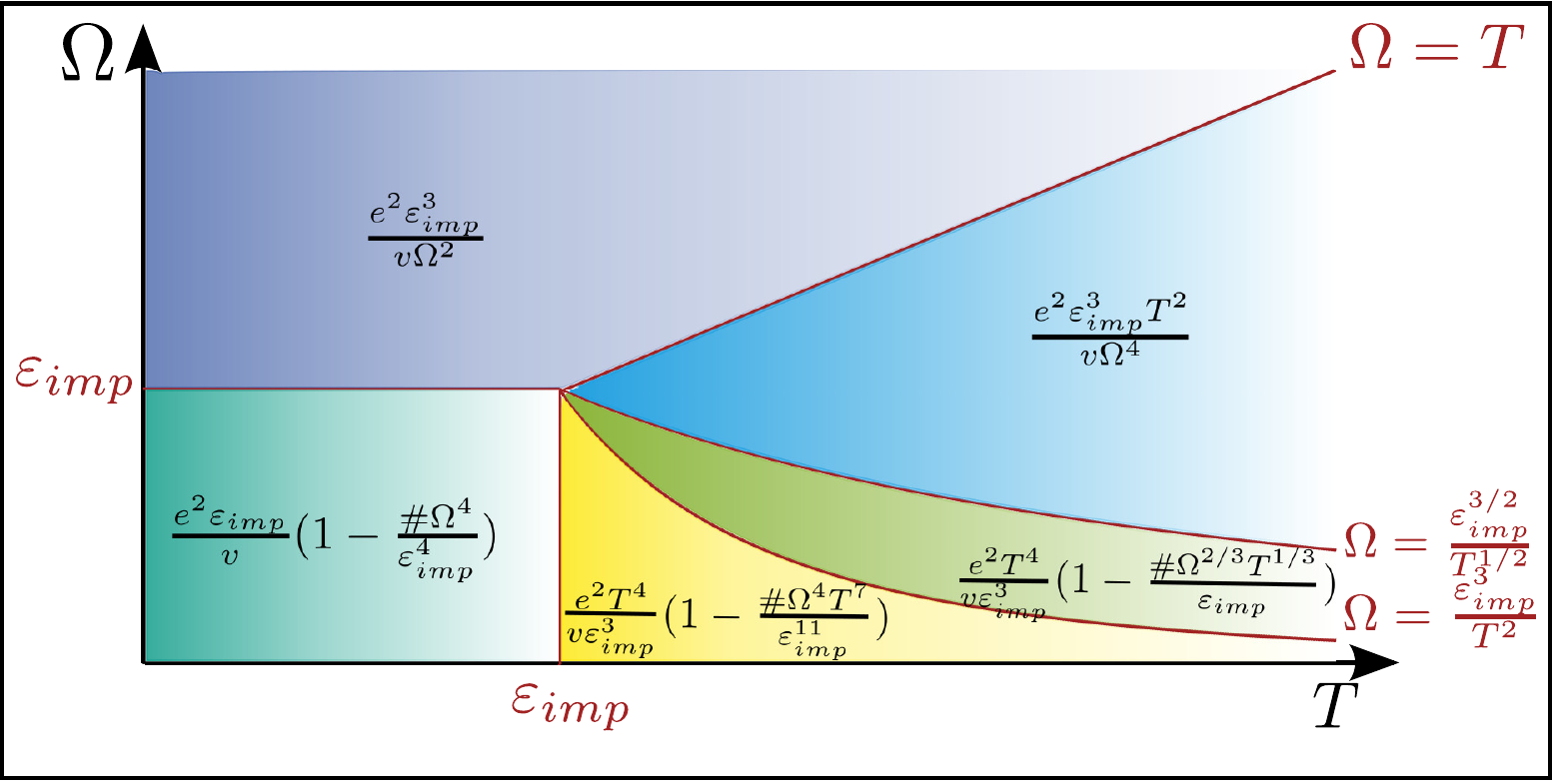}
\caption{Behavior of the conductivity $\sigma_{xx}$ in a Weyl semimetal with Coulomb impurities in dependence of temperature and magnetic field. Scaling of dominant contribution to the conductivity in each of the parameter regimes is shown (the Fermi velocity $v$ is set to unity).
Equations describing borderlines between the regimes are also indicated.
The symbol $\#$ denotes numerical coefficients.}
\label{phasediagrammCoul}
\end{center}
\end{figure*}

We turn now to the domain $T>\Omega$ characterized by a temperature-dependent screening.
Here contributions of higher LLs are important. We use the semiclassical expression (\ref{Drude}) for the
conductivity together with Eq.~(\ref{cyclotron}) and
\begin{equation}
\frac{1}{\tau(\varepsilon)}\sim \frac{\gamma(T) \varepsilon^2}{v^3} \sim \frac{\varepsilon_\text{imp}^3\varepsilon^2}{T^4}.
\end{equation}
A closer look reveals a necessity to distinguish between several regimes.
For $\varepsilon<\Omega^{2/3}T^{4/3}/\varepsilon_\text{imp}$, one can neglect the unity in the denominator of the integrand in Eq.~(\ref{Drude}).
Therefore, for $T>\varepsilon_\text{imp}^3/\Omega^2$,
the conductivity of a system with screened Coulomb impurities takes the form
\begin{align}\label{intermedconcol}
\sigma_{xx}&\sim \frac{e^2 T^2 \varepsilon_\text{imp}^3}{v \Omega^4} \sim \frac{e^2 T^2 N_\text{imp}v^2}{\Omega^4} \propto \frac{1}{H^2}.
\end{align}

For $T<\varepsilon_\text{imp}^3/\Omega^2$ the conductivity is dominated
by the range of $\varepsilon$ where we can neglect the magnetic field:
\begin{equation}
\sigma_{xx}\sim \frac{e^2 v^2}{\gamma(T)}\sim \frac{e^2 T^4}{v \varepsilon_\text{imp}^3} \sim \frac{e^2 T^4}{v^4 N_\text{imp}}.
\end{equation}
The $H$-dependent correction to this result turns out to be different in the two subregimes.
The first subregime is defined by $\varepsilon_{\text{imp}}^3/T^2<\Omega<\varepsilon_{\text{imp}}^{3/2}/T^{1/2}$.
The result is obtained similarly to Eq.~(\ref{correction}):
\begin{eqnarray}
\delta\sigma_{xx}\sim  -\frac{e^2 v^2}{\gamma(T)}\frac{\Omega^{2/3} T^{1/3}}{\varepsilon_\text{imp}} \sim
- \frac{e^2 T^{13/3}}{v^5 N_\text{imp}^{4/3}} \Omega^{2/3} \propto - H^{1/3}.
\nonumber
\\
\label{correction-1}
\end{eqnarray}
For weaker magnetic fields, $\Omega<\varepsilon_{\text{imp}}^3/T^2$, the energies
dominating the magnetic-field dependence of the conductivity are below $\varepsilon_\text{imp}$.
As a consequence, one should replace $\varepsilon$ by $\varepsilon_{\text{imp}}$ in the parameter $\omega_c\tau$:
\begin{equation}
\omega_c\tau\sim\frac{\Omega^4T^8}{\varepsilon_\text{imp}^{12}}.
\end{equation}
The $H$-dependent correction to the conductivity for $\Omega<\varepsilon_\text{imp}^{3/2}/T^{1/2}$ is
\begin{eqnarray}
\delta\sigma_{xx}\sim  -\frac{e^2 \Omega^{4}T^{11} }{v \varepsilon_\text{imp}^{14}} \propto - H^{2}.
\label{correction-1.1}
\end{eqnarray}
We thus see that the non-analytic ($H^{1/3}$) magnetoresistance does not survive the limit $H\to 0$
for the case of charged impurities, in contrast to the case of white-noise disorder.

For the lowest magnetic fields $\Omega<\varepsilon_\text{imp}$ and temperatures $T<\varepsilon_\text{imp}$,
all Landau levels overlap.
We thus expect that a zero-$H$ calculation, as recently carried
out in Refs.~\onlinecite{PhysRevB.91.035202} and \onlinecite{PhysRevB.91.195107}, should be applicable.
In our notations, the result of those papers is written as
\begin{equation}
\sigma_{xx}\sim e^2 N^{1/3}_\text{imp}
\sim \frac{e^2 \varepsilon_\text{imp}}{v}.
\label{sigma_imp}
\end{equation}
In this regime, disorder is strong. This means that the $H$-dependent correction to the conductivity can be calculated under the assumption that $1/\tau\sim  \Gamma\sim\varepsilon_\text{imp}$ and $\omega_c\sim\Omega^2/\Gamma$. The $H$-dependent correction
to Eq.~(\ref{sigma_imp}) for $\Omega\ll T \ll \varepsilon_\text{imp}$ is calculated similarly to Eq.~(\ref{Drude-strong-2})
and takes the form:
 \begin{eqnarray}
\delta\sigma_{xx}\sim  -\frac{e^2 \Omega^{4} }{v \varepsilon_\text{imp}^{3}} \propto - H^{2}.
\label{correction-2}
\end{eqnarray}
Finally, in the regime $T\ll \Omega \ll \varepsilon_\text{imp}$ an analogous consideration with $1/\tau\sim \Omega^2/\varepsilon_\text{imp}$
yields no $H$ dependent correction to the leading order.

The conductivity in all regimes is summarized in Fig.~\ref{phasediagrammCoul}. For each regime, the scaling of the dominate contributions is shown. We can summarize the results according to the parameter $T/\varepsilon_{\text{imp}}$, yielding to
\begin{equation}
\sigma_{xx}\sim
\left\{
  \begin{array}{ll}
    \dfrac{e^2 \varepsilon_{\text{imp}}^3}{v\Omega^2}\propto\dfrac{1}{H}, &\quad  \Omega \gg T, \\[0.3cm]
    \dfrac{e^2\varepsilon_{\text{imp}}}{v}\left(1-\dfrac{\Omega^4}{\varepsilon_{\text{imp}}^4}\right), &\quad\Omega \ll T
  \end{array}
\right.
\label{coul-sigma-low}
\end{equation}
for $T/\varepsilon_{\text{imp}}<1$. In the opposite limit, we need to distinguish between more regimes
\begin{equation}
\sigma_{xx}\!\sim\!
\left\{
  \begin{array}{ll}
    \dfrac{e^2 \varepsilon_{\text{imp}}^3}{v\Omega^2}\propto\dfrac{1}{H}, &\quad\!\!\!\!  \Omega \gg T, \\[0.3cm]
    \dfrac{e^2\varepsilon_{\text{imp}}^3T^2}{v\Omega^4}, &\quad\!\!\!\!\dfrac{\varepsilon_{\text{imp}}^{3/2}}{T^{1/2}}\ll\Omega \ll T, \\[0.3cm]
    \dfrac{e^2T^4}{v\varepsilon_{\text{imp}}^3}\!\left(1-\dfrac{\#\Omega^{2/3}T^{1/3}}{\varepsilon_{\text{imp}}}\right), &\quad\!\!\!\!\dfrac{\varepsilon_{\text{imp}}^{3}}{T^{2}}\!\ll\Omega \ll\!\dfrac{\varepsilon_{\text{imp}}^{3/2}}{T^{1/2}}, \\[0.3cm]
    \dfrac{e^2T^4}{v\varepsilon_{\text{imp}}^3}\!\left(1-\dfrac{\#\Omega^4T^7}{\varepsilon^{11}_{\text{imp}}}\right), &\quad\!\!\!\!\Omega\ll\dfrac{\varepsilon_{\text{imp}}^{3}}{T^{2}}.
  \end{array}
\right.
\label{coul-sigma-high}
\end{equation}
Here and in Fig.~\ref{phasediagrammCoul}, the symbol $\#$ denotes numerical coefficients.
We observe that the limits $T\to0$ and $H\to0$ are interchangeable for Coulomb impurities.

\section{Magnetoresistance}\label{sec:magnetoresistance}

Using the results of the previous sections, it is straightforward to evaluate the magnetoresistance described by
\begin{equation}\label{magres}
\Delta_\rho(H)=\frac{\rho_{xx}(H)-\rho_{xx}(0)}{\rho_{xx}(0)}.
\end{equation}
The resistivity is given by $\rho_{xx}=\sigma_{xx}/(\sigma^2_{xx}+\sigma_{xy}^2)$.
Since we considered the case of zero chemical potential $\mu=0$, the Hall conductivity is zero and we can calculate the magnetoresistance with
\begin{equation}
\Delta_\rho(H)=\frac{\sigma_{xx}(0)}{\sigma_{xx}(H)}-1.
\end{equation}
In the following, we calculate this relative magnetoresistance for both models of disorder, pointlike impurities and charged impurities.

\subsection{Pointlike Impurities}

\begin{figure}
\begin{center}
\includegraphics[scale=0.6]{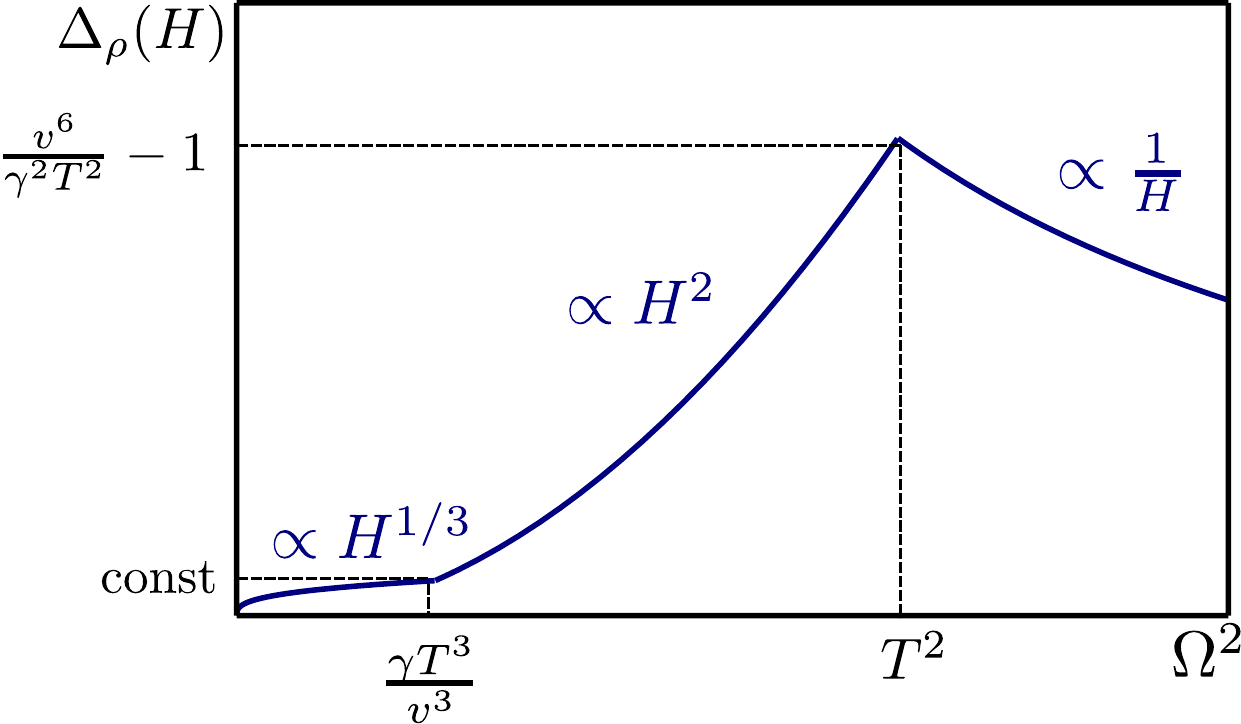}
\end{center}
\caption{Dependence of the magnetoresistance on magnetic field ($\Omega^2\propto H$) for the case of white-noise disorder (weak disorder regime). Magnetoresistance in the range of relatively weak magnetic fields, $\Omega\ll v^3/\gamma$, is shown. For larger fields the magnetoresistivity would vanish.
Scaling of the magnetoresistance in different regimes, regimes boundaries and the corresponding values of $\Delta\rho$ are indicated.}
\label{fig:magres}
\end{figure}

We fix the value of $T\gamma$ and analyze the evolution of the magnetoresistance with increasing magnetic field (which corresponds to a vertical cross-section in Fig.~\ref{phasediagramm}).
Using Eq.~\eqref{short-range-sigma}, we obtain the magnetoresistance
\begin{equation}
\Delta_\rho=
\left\{
  \begin{array}{ll}
    \dfrac{\pi^{4/3}}{12^{2/3}}\dfrac{ v \Omega^{2/3}}{\gamma^{1/3}T}, &\quad  \Omega \ll \dfrac{\gamma^{1/2}T^{3/2}}{v^{3/2}}, \\[0.3cm]
    \dfrac{189}{124 \pi^4}\dfrac{v^6 \Omega^4}{\gamma^2 T^6}-1, &\quad
\dfrac{\gamma^{1/2}T^{3/2}}{v^{3/2}} \ll \Omega \ll T , \\[0.3cm]
    8\pi^2\dfrac{v^6 }{\gamma^2 \Omega^2}-1, &\quad  T \ll \Omega \ll \dfrac{v^3}{\gamma}.
  \end{array}
\right.
\label{short-rane-deltarho}
\end{equation}
For $\Omega>v^3/\gamma$,
the magnetoresistance vanishes, because a further increase of magnetic field leads again to a regime of overlapping LL.
The magnetoresistance in different regimes is visualized in Fig.~\ref{fig:magres}.

It is worth emphasizing that the vanishing of the density of states of Weyl semimetal at $\varepsilon=0$ in zero $H$
translates into the non-analytic, $H^{1/3}$, behavior of the magnetoresistance for weak pointlike impurities, which
persists down to $H=0$. This should be contrasted with the $H^{1/2}$-magnetoresistance in graphene found in
Ref. \onlinecite{PhysRevB.87.165432}. The square-root magnetoresistance in graphene does not actually
survive the limit $H\to 0$, since even weak white-noise scalar disorder is marginally relevant in graphene and establishes
a finite density of states at the Dirac point. As a result, the true $H\to 0$ asymptotics of the magnetoresistance in
graphene is parabolic (although the crossover between $H^2$ and $H^{1/2}$ may occur at very weak magnetic fields).
By contrast, weak white-noise disorder in 3D systems with linear spectrum is not capable of establishing a finite DOS
at $\varepsilon=0$. This leads to the remarkable observation of the non-analyticity of the transversal magnetoresistance.
This non-analyticity can be considered as a finite-$H$ counterpart of the non-commutativity of the limits $T\to 0$ and
$\omega\to 0$ that takes place in zero magnetic field.

We also stress that in the limit of low temperatures $T\to 0$,
the magnetoresistance for pointlike impurities is very sharp, with the maximum at $H\propto T^2 \to 0$ growing
as $\Delta_\rho \propto 1/T^{2}\to \infty$. This is a manifestation of the non-commutativity of the limits $T\to 0$ and $H\to 0$
discussed at the end of Sec. \ref{cond-highT}.

\subsection{Charged Impurities}

For the case of charged impurities, we perform the analysis in a similar way.
Specifically, we fix the parameter $T/\varepsilon_\text{imp}$ and then consider the magnetoresistivity when the magnetic field is swept.

\begin{figure}
\includegraphics[width=7cm]{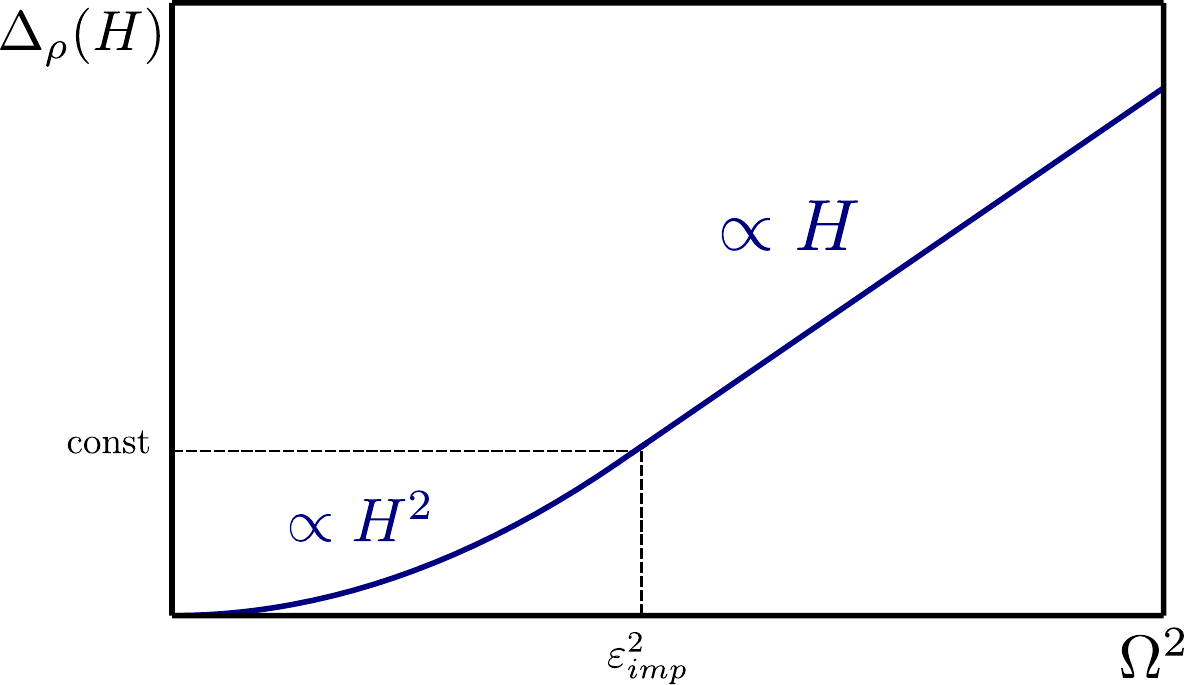}
\caption{Magnetoresistivity for a Weyl semimetal with Coulomb impurities in the limit of low temperatures, $T<\varepsilon_\text{imp}$.}
\label{magres-coul-lowT}
\end{figure}

For $T/\varepsilon_{\text{imp}}<1$, Eq.~\eqref{coul-sigma-low} is used to obtain the magnetoresistivity, reading
\begin{equation}
\Delta_\rho\sim
\left\{
  \begin{array}{ll}
    \dfrac{\Omega^{4}}{\varepsilon_\text{imp}}, &\quad  \Omega \ll \varepsilon_\text{imp}, \\[0.3cm]
    \dfrac{\Omega^2}{\varepsilon_\text{imp}}-1, &\quad
\varepsilon_\text{imp} \ll \Omega.
  \end{array}
\right.
\label{Coul-low-deltarho}
\end{equation}
This type of behavior in the high field limit was identified in an early work by Abrikosov, Ref.~\onlinecite{PhysRevB.58.2788}. For lower magnetic fields, the LLs overlap, which implies that the conductivity is essentially the same as in the absence of magnetic field. Therefore, the magnetoresistance vanishes quadratically, as visualized in Fig.~\ref{magres-coul-lowT}.

\begin{figure}
\begin{center}
\includegraphics[scale=0.6]{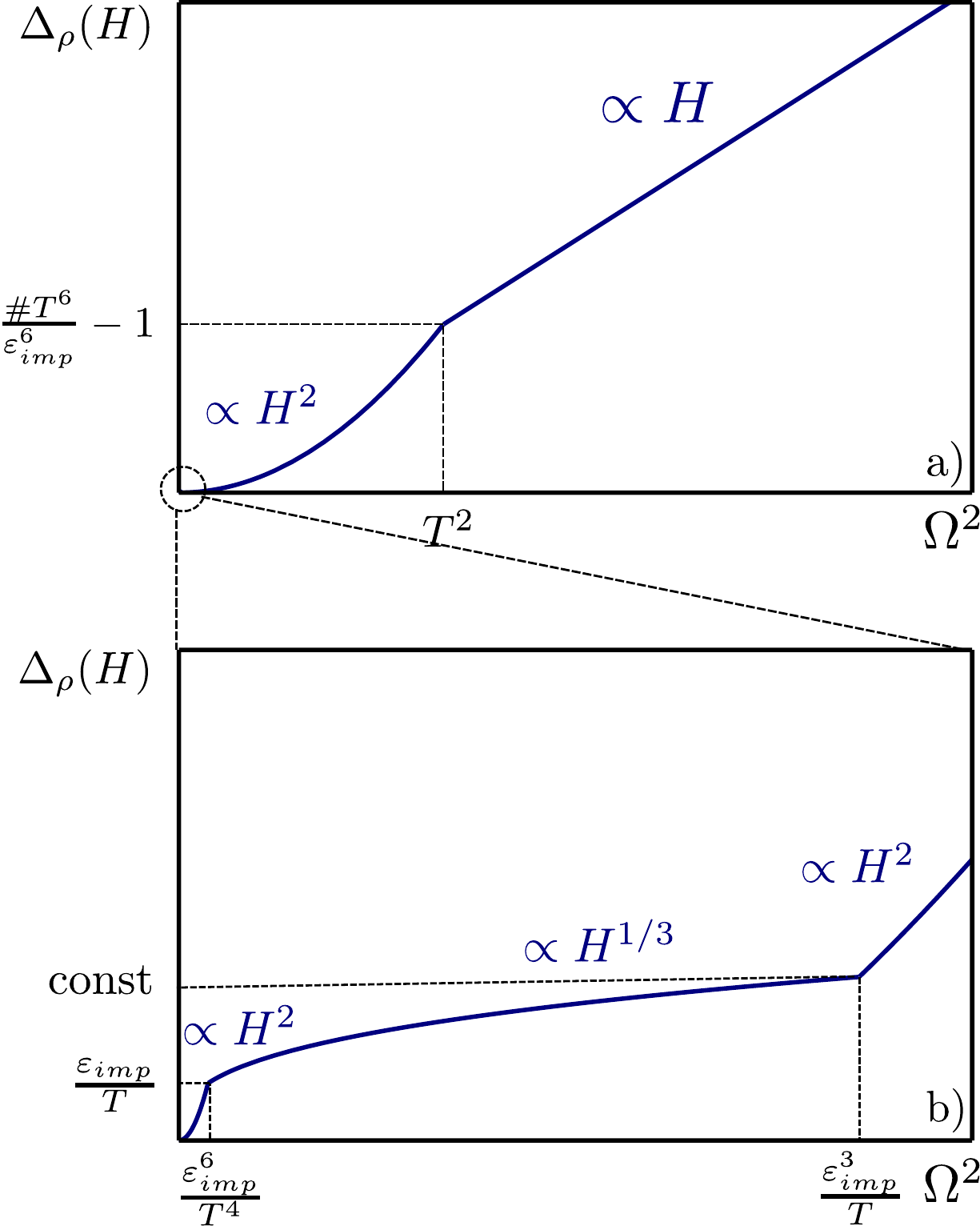}
\end{center}
\caption{ Magnetoresistance of a Weyl semimetal with Coulomb impurities for sufficiently high temperatures, $T>\varepsilon_{\text{imp}}$, in the whole range of magnetic fields (a). The low-field region is magnified in panel (b).
}
\label{fig:magrescoulhigh}
\end{figure}

In the temperature limit $T/\varepsilon_\text{imp}>1$, we use Eq.~\eqref{coul-sigma-high} to obtain the magnetoresistivity. This leads to
\begin{equation}
\Delta_\rho\sim
\left\{
  \begin{array}{ll}
    \dfrac{\Omega^{4}T^7}{\varepsilon_\text{imp}^{11}}, &\quad  \Omega \ll \dfrac{\varepsilon_\text{imp}^3}{T^2}, \\[0.3cm]
    \dfrac{\Omega^{2/3}T^{1/3}}{\varepsilon_\text{imp}}, &\quad \dfrac{\varepsilon_\text{imp}^3}{T^2}\ll\Omega\ll\dfrac{\varepsilon_\text{imp}^{3/2}}{T^{1/2}},\\[0.3cm]
    \dfrac{T^2\Omega^4}{\varepsilon_\text{imp}^6}-1, &\quad \dfrac{\varepsilon_\text{imp}^{3/2}}{T^{1/2}}\ll\Omega\ll T,\\[0.3cm]
    \dfrac{\Omega^2T^4}{\varepsilon_\text{imp}^6}-1, &\quad T \ll \Omega.
  \end{array}
\right.
\label{Coul-high-deltarho}
\end{equation}
In the high-field limit, the magnetoresistivity is again linear in magnetic field. The low-field limit is dominated
by a quadratic dependence on magnetic field. This is visualized in Fig.~\ref{fig:magrescoulhigh}.

\subsection{Comparison to experiment}

As has been mentioned in the Introduction, two recent works \cite{RIS_1,2014arXiv1405.6611F}  reported a strong, approximately linear magnetoresistance in the Dirac semimetal Cd$_{3}$As$_{2}$ in the range of strong magnetic fields.
We have found that a model of pointlike impurities yields a very different behavior of the magnetoresistance
and thus cannot explain the experiment. On the other hand, a more realistic model of disorder---Coulomb impurities---does
lead to a linear magnetoresistance in the region of strong magnetic fields,
yielding a tentative explanation of spectacular experimental observations.
In particular,
the experiment of Ref.~\onlinecite{RIS_1} shows a linear longitudinal resistivity starting
at the magnetic field $H=1.1\text{T}$ for $T=300\text{K}$.
This is in a good agreement with the onset of linear magnetoresistance at $\Omega\agt T$ [Eqs.~\eqref{Coul-low-deltarho} and \eqref{Coul-high-deltarho}]
which corresponds to $H\sim 1\text{T}$ for $T=300\text{K}$. At the same time, in the range of weaker magnetic field, $H<1\text{T}$ the experimentally measured magnetoresistance is parabolic, which again agrees with our result for Coulomb impurities. The slope $0.4\text{m}\Omega\text{cm}$ of the linear resistivity in experiment \cite{RIS_1} agrees with our estimate $\sim 0.3\text{m}\Omega\text{cm}$ obtained under the assumption that the carrier concentration is of the order of density of Coulomb impurities.
Furthermore, the magnetoresistance in Ref.~\onlinecite{0953-8984-27-15-152201} shows two inflection points similar to our result of the
magnetoresistance visualized in Fig.~\ref{fig:magrescoulhigh} (b).

Let us point out that we have performed the analysis for the neutrality point, $\mu=0$, while the experiment is carried out at a finite carrier density. The latter leads to  an observation of Shubnikov-de Haas oscillations which are absent in our $\mu=0$ theory.
We expect that our theory should become applicable to a system with finite carrier density for sufficiently strong magnetic field, such that the chemical potential is located between the zeroth and the first LL.

\section{Summary and discussion}\label{sec:summary}

To summarize, we have developed a theory of the magnetoresistivity of Weyl and Dirac semimetals at the neutrality point.
We have considered  two alternative models of disorder: (i) short-range impurities and (ii) charged (Coulomb) impurities and treated the  impurity in the framework of  the self-consistent Born approximation. We have found that an unusual broadening of Landau levels leads to a variety of regimes of the resistivity scaling in the temperature-magnetic field plane.

The behavior of the magnetoresitance is essentially different for two types of disorder, with the difference originating from the dependence of screening
of charged impurities on magnetic field. In the limits of strongest magnetic fields $H$, the magnetoresistivity vanishes as $1/H$ for pointlike impurities, Eq.~\eqref{short-rane-deltarho}, while it is linear and positive in the model with Coulomb impurities, Eqs.~\eqref{Coul-low-deltarho} and \eqref{Coul-high-deltarho}, in agreement with experimental observations in Refs. \onlinecite{RIS_1} and \onlinecite{2014arXiv1405.6611F}.
The prefactor of this linear magnetoresistance is approximately temperature-independent for low temperatures.
In the low-field limit, we find a quadratic magnetoresistivity for screened Coulomb impurities,
Eq.~\eqref{Coul-low-deltarho}.
By contrast, the low-field magnetoresistance for pointlike impurities shows a non-analytic behavior, $\propto H^{1/3}$.

It should be emphasized, however, that our theory was developed for the Dirac point, $\mu=0$, whereas the experiments have been carried out at a finite carrier density which manifested itself in a non-zero Hall resistivity and in Shubnikov-de-Haas oscillations of the resistivity. For non-zero $\mu$ the present theory is justified only in the range of sufficiently high temperatures or magnetic fields, $\mu<\text{max}[\Omega,T]$, for which the chemical potential is located between the zeroth and first Landau levels.

A generalization of the theory on the case of finite $\mu$ and lower magnetic fields remains a prospect for future research. Another interesting generalization of our theory would be the analysis of the effect of magnetic field on the critical region around the
transition ($\gamma=\gamma_c$) between weak and strong disorder.

Further, our theory assumed the absence of internodal scattering between different Weyl points and diagonal scalar impurity potential.
The internodal scattering is believed to be weaker than the intranodal one in realistic Weyl semimetals. However, even weak internodal
scattering is crucially important for quantum effects, especially for those related to the famous chiral anomaly.
Since the transverse magnetoresistance (in contrast to the longitudinal magnetoresistance) is not directly affected by the chiral anomaly,
we expect that our semiclassical calculation based on the SCBA will not change qualitatively in the presence of internodal scattering,
provided the two Weyl points correspond to the same energy. In the case of non-degenerate Weyl points, the internodal scattering is expected
to establish a finite density of states at any energy even in the limit of zero magnetic field. In this situation, we expect that the
non-analytic, $H^{1/3}$, behaviour of the magnetoresistivity at $H\to 0$ for weak pointlike impurities (stemming from the vanishing DoS)
would be smeared at the lowest $H$.
A detailed study of this case is a prospect for future work. In a similar way, the nondiagonal matrix elements of disorder potential should not
affect our SCBA results qualitatively as long as they do not induce a finite DoS.

Finally, the Coulomb interaction between quasiparticles and the particle-hole recombination in a finite geometry
 are additional sources of the magnetoresistance at the charge neutrality point,
 similarly to 2D compensated systems, see Refs.~\onlinecite{PhysRevLett.114.156601}, \onlinecite{PhysRevB.91.035414} and references therein.
 These effects are expected to be important for the description of the high-temperature magnetoresistance.

\textit{Note added.} After the completion of our paper, we became aware of
very recent related works on magnetotransport in Weyl semimetals.
In Ref.~51 the magnetoresistance is analyzed for a model with smooth random
potential. In Ref.~52, a transversal magnetoresistance in Weyl semimetals 
with Coulomb impurities is calculated for a small ``fine-structure constant'' [cf. Eq.~(\ref{fine})],
and the result is consistent with our results obtained for $e^2/v\gtrsim 1$. In Ref.~53, the role of
Fermi arcs in a finite-width slab of a Weyl semimetal subject to a static
magnetic field is addressed in the context of non-local response. While our
calculations have been performed for macroscopic samples, it would be interesting
to investigate a possible effect of Fermi arcs on the transversal
magnetoresistance in a thin-slab geometry.

\section*{Acknowledgments}
We acknowledge useful discussions with
U. Briskot, V. Kachorovskii, A. Levchenko, P. Ostrovsky, J. Schmalian, and B. Yan.
The work was supported by EU Network FP7-PEOPLE-2013-IRSES (project ``InterNoM''),
by the Priority Programme 1666 ``Topological Insulators''
of the Deutsche Forschungsgemeinschaft (DFG-SPP 1666),
and by German-Israeli Foundation (GIF).

\appendix
\section{Shape of the broadening of separated Landau levels}
\label{app:shape}

In this appendix we analyze the shape of the LL broadening at $\varepsilon<\varepsilon_*=\Omega(\Omega/A)^{1/5}$ (when LL are well separated).
The maximum of $\Gamma(\varepsilon)$ around $\varepsilon\sim W_N$ is located at $\varepsilon\simeq W_N+\Gamma(W_N)/2^{2/3}$ and
is given by $\Gamma_\text{top} = \Gamma(W_N)3^{1/2}/2^{2/3}$, where $\Gamma(W_N)=(A/2)^{2/3}W_N^{1/3}.$
For brevity, we use the abbreviation $\Gamma_N=\Gamma(W_N)$.

For $\varepsilon>W_N+\Gamma_N$, the peak in $\Gamma(\varepsilon)$ decreases as
\begin{equation}
\Gamma(\varepsilon)\simeq \Gamma_N \sqrt{\frac{2 \Gamma_N}{\varepsilon-W_N}} = \frac{A}{\sqrt{2}} \sqrt{\frac{W_N}{\varepsilon-W_N}},
\label{right-tail}
\end{equation}
and reaches the value of the background at $\varepsilon\sim W_N+\Omega (\Omega/\varepsilon)^3$.
This value is always smaller than $W_{N+1}\simeq W_N+\Omega^2/(2\varepsilon)$. Thus, in the range
$W_N+\Omega (\Omega/\varepsilon)^3<\varepsilon<W_{N+1}-\Gamma(W_{N+1})$, the Landau level broadening is
of the order of $A(\varepsilon/\Omega)^2$ (zero-$H$ result).

\begin{figure}
\centerline{\includegraphics[width=8cm]{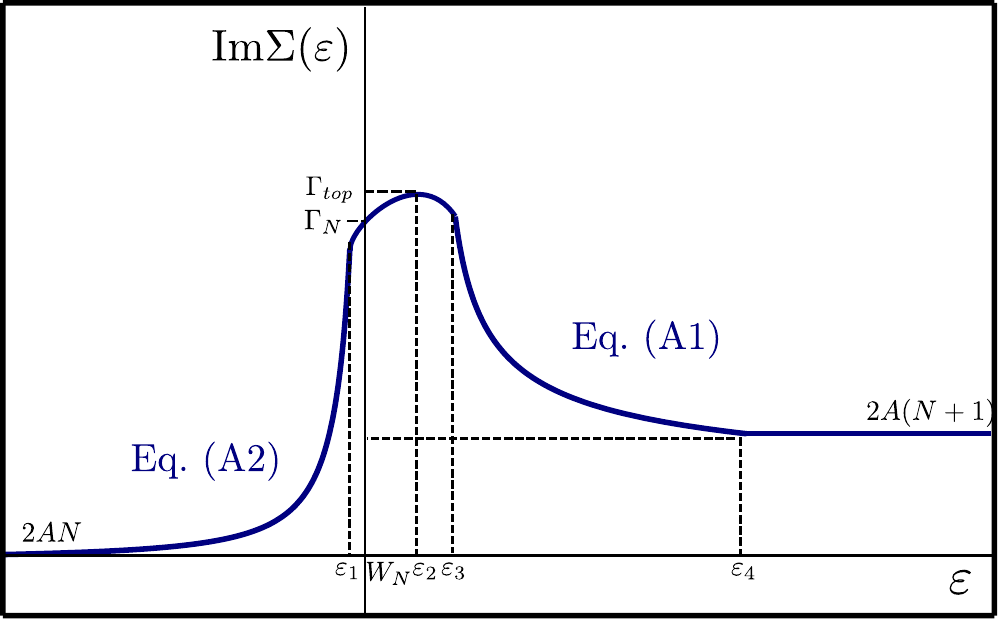}}
\caption{Schematic illustration of the Landau-level broadening.
At energies $\varepsilon<\varepsilon_1$, such that $|\varepsilon_1-W_N|\sim \Gamma_N$, the broadening is given by Eq.~(\ref{left-side}).
The maximum of $\text{Im}\Sigma$ is $\Gamma_{\text{top}} \simeq \Gamma_N 3^{1/2}/2^{2/3} \simeq 1.1 \Gamma_N$ and is achieved at $\varepsilon_2=W_N+\Gamma_N/2^{2/3}\simeq W_N+0.63 \Gamma_N$. At $\varepsilon_3\sim W_N+\Gamma_N$ the tail (\ref{right-tail}) develops. At $\varepsilon_4\sim W_N+\Omega (\Omega/\varepsilon)^3$ this tail
reaches the background $2A(N+1)$.}
\label{fig3}
\end{figure}

On the left side of the peak, for $\varepsilon<W_N-\Gamma_N$, the solution of the self-consistency equation (\ref{SCEq})
yields
\begin{equation}
\Gamma(\varepsilon)\simeq 2A\frac{\varepsilon^2}{\Omega^2}
\left[1+\frac{A\sqrt{W_N}}{[2(W_N-\varepsilon)]^{3/2}-A\sqrt{W_N}}\right].
\label{left-side}
\end{equation}
which matches $\Gamma_N$ at $\varepsilon \sim W_N-\Gamma_N-A\varepsilon^2/\Omega^2$.
The decrease of left side of the peak from $\Gamma(\varepsilon)\sim \Gamma_N$ to $\Gamma(\varepsilon)\sim A \varepsilon^2/\Omega^2$ is thus
very sharp.
The Landau-level broadening for the case of well separated levels is shown schematically in Fig.~\ref{fig3}.

\section{Vertex corrections}\label{app:vertcorr}

In this appendix, we analyze the vertex corrections in the diagram for the conductivity.
We calculate the following diagram that describes the dressing of a current vertex by disorder lines:
\begin{align}
\vcenter{\hbox{\includegraphics[scale=0.3]{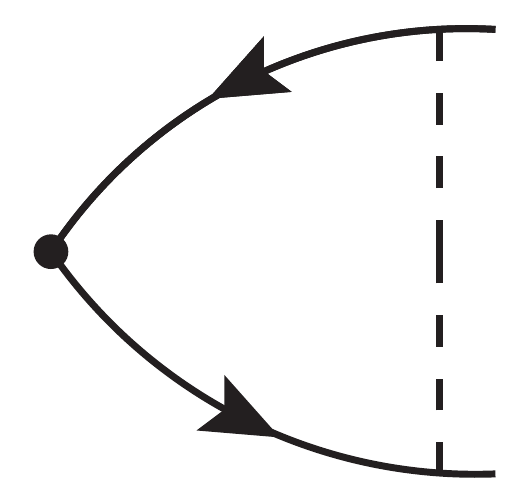}}}=&ev\gamma\int\frac{dp_y}{2\pi}\int\frac{dp_z}{2\pi}\int dx\int dx'\delta(x-x')\nonumber\\
&\times
\hat{G}^A(\varepsilon, p_y, p_z, x, x')\sigma_x\hat{G}^R(\varepsilon, p_y, p_z, x', x).
\nonumber
\\
\end{align}
The diagonal elements for the resulting matrix vanish after the integration over $x$, $x'$ and $p_y$ because of the orthogonality of the Hermite functions. The off-diagonal elements $v_{x,12}$ are equal. They read
\begin{widetext}
\begin{align}\label{vertcorr}
v_{x,12}=ev\int\frac{dp_z}{2\pi}\frac{\gamma eH}{2\pi c}\sum_{n=0}^{N_\text{max}}\frac{\varepsilon-\Sigma_2^R+vp_z}{(\varepsilon-\Sigma_1^R-vp_z)(\varepsilon-\Sigma_2^R+vp_z)+\Omega^2n}\frac{\varepsilon-\Sigma_1^A-vp_z}{(\varepsilon-\Sigma_1^A-vp_z)(\varepsilon-\Sigma_2^A+vp_z)+\Omega^2(n+1)}.
\end{align}
For energies away from the Weyl point, the difference between the two self-energies, $\Sigma_1$ and $\Sigma_2$, can be neglected.
Furthermore, for high Landau levels we can express the Landau level index $n$ in terms of the momenta in $p_x$ and $p_y$ direction.
The correction to the current vertex then simplifies as follows
\begin{align}
v_{x,12}=&ev\gamma\int \frac{dp^3}{(2\pi)^3}
\frac{(\varepsilon-\Sigma^A)(\varepsilon-\Sigma^R)-v^2p_z^2}{((\varepsilon-\Sigma^A)^2-v^2p^2)((\varepsilon-\Sigma^R)^2-v^2p^2-\Omega^2)},
\end{align}
which can be cast in the form:
\begin{align}
v_{x,12}=&\frac{(\varepsilon-\Sigma^A)(\varepsilon-\Sigma^R)}{(\varepsilon-\Sigma^R)^2-(\varepsilon-\Sigma^A)^2+\Omega^2}\frac{ev\gamma}{\pi}
\int\frac{dp}{2\pi}p^2\left[\frac{1}{(\varepsilon-\Sigma^A)^2-v^2p^2-\Omega^2}-\frac{1}{(\varepsilon-\Sigma^R)^2-v^2p^2}\right]
\nonumber\\&+\frac{ev\gamma}{6\pi}\int\frac{dp}{2\pi}p^2\left[\frac{1}{(\varepsilon-\Sigma^A)^2-v^2p^2-\Omega^2}+\frac{1}{(\varepsilon-\Sigma^R)^2-v^2p^2}\right]\nonumber\\&-\frac{(\varepsilon-\Sigma^A)^2+(\varepsilon-\Sigma^R)^2-\Omega^2}{(\varepsilon-\Sigma^R)^2-(\varepsilon-\Sigma^A)^2+\Omega^2}\frac{ev\gamma}{6\pi}\int\frac{dp}{2\pi}p^2\left[\frac{1}{(\varepsilon-\Sigma^A)^2-v^2p^2-\Omega^2}-\frac{1}{(\varepsilon-\Sigma^R)^2-v^2p^2}\right].
\end{align}
The integrals over momenta here can be identified with those for the self-energy.
The shift in the denominator by $\Omega^2$ of the integrands is unimportant (similarly to the difference
between $\Sigma_1$ and $\Sigma_2$) and can be neglected; it is then
sufficient to keep $\Omega^2$ in the prefactors of the integrals.
The corrections to the current in $x$-direction then reads
\begin{align}
v_{x,12}=&ev\left(\frac{(\varepsilon-\Sigma^A)(\varepsilon-\Sigma^R)}{(\varepsilon-\Sigma^R)^2-(\varepsilon-\Sigma^A)^2+\Omega^2}\left[\frac{\Sigma^A}{\varepsilon-\Sigma^A}-\frac{\Sigma^R}{\varepsilon-\Sigma^R}\right]+\frac{1}{6}\left[\frac{\Sigma^A}{\varepsilon-\Sigma^A}+\frac{\Sigma^R}{\varepsilon-\Sigma^R}\right]\right.\nonumber\\&-\left.\frac{1}{6}\frac{(\varepsilon-\Sigma^A)^2+(\varepsilon-\Sigma^R)^2-\Omega^2}{(\varepsilon-\Sigma^R)^2-(\varepsilon-\Sigma^A)^2+\Omega^2}\left[\frac{\Sigma^A}{\varepsilon-\Sigma^A}-\frac{\Sigma^R}{\varepsilon-\Sigma^R}\right]\right).
\end{align}
For weak disorder, using $\Sigma^{R,A}\ll \varepsilon$, after some algebra we arrive at:
\begin{align}
v_{x,12}=\frac{ev}{3}\frac{4i\varepsilon\Gamma}{4i\varepsilon\Gamma+\Omega^2}.
\end{align}
This result matches to the result in Ref.~\onlinecite{PhysRevB.89.014205} for the case $H=0$.
The summation over the ladder of impurity lines yields
\begin{align}
V^\text{tr}(\varepsilon)=\left[1-\frac{1}{3}\frac{4i\varepsilon\Gamma(\varepsilon)}{4i\varepsilon\Gamma(\varepsilon)+\Omega^2}\right]^{-1},
\end{align}
leading to Eq.~(\ref{Vtr}) and hence
\begin{align}
\tau^\text{tr}(\varepsilon)=\frac{3}{2}\tau^q(\varepsilon)
\end{align}
in the Drude formula \eqref{Drude}.

This result is not applicable at the Weyl point, where $\Sigma_1\neq\Sigma_2$. To simplify Eq.~\eqref{vertcorr}, we can use $\Sigma_2=0$ which is justified for weak disorder. Equation \eqref{vertcorr} simplifies then as follows
\begin{align}\label{vertcorr1}
v_{x,12}=ev\int\frac{dp_z}{2\pi}\frac{\gamma eH}{2\pi c}\sum_{n=0}^{N_\text{max}}
\frac{\varepsilon+vp_z}{(\varepsilon-\Sigma_1^R-vp_z)(\varepsilon+vp_z)+\Omega^2n}\frac{\varepsilon-\Sigma_1^A-vp_z}
{(\varepsilon-\Sigma_1^A-vp_z)(\varepsilon+vp_z)+\Omega^2(n+1)}.
\end{align}
We find that this integral is proportional to $A/\Omega$ and hence the vertex corrections are small.
This is in agreement with the neglect of vertex corrections in the calculation of the strong-$H$ conductivity dominated by
the lowest Landau level in Ref.~\onlinecite{PhysRevB.58.2788}.

\section{Details of calculation of the conductivity for high temperatures}
\label{app:highT}
In this Appendix, we present details of the evaluation of the conductivity in the regime of high temperatures
when many Landau levels contribute to the result.
We first calculate the conductivity without current-vertex corrections and include the vertex corrections in the end of the
calculation. Starting from Eqs.~\eqref{sigmaQ} and \eqref{Qne},
we represent $Q_n(\varepsilon)$ with $V^\text{tr}=1$ as a sum of the two terms related to $\text{Im}G_{11}^R$ and $\text{Im}G_{22}^R$ in Eq.~\eqref{Qne},
respectively:
\begin{eqnarray}
Q_n(\varepsilon)&=&\Gamma^2 \int_{-\infty}^\infty \frac{dz}{2\pi} \
\frac{(\varepsilon^2+z^2+\Omega^2 n +\Gamma^2+2 \varepsilon z)(\varepsilon^2+z^2+\Omega^2 (n+1)+\Gamma^2+2 \varepsilon z)-4\varepsilon^2z^2}
{[(\varepsilon^2-z^2-\Omega^2 n -\Gamma^2)^2+4\varepsilon^2\Gamma^2][(\varepsilon^2-z^2-\Omega^2 (n+1)-\Gamma^2)^2+4\varepsilon^2\Gamma^2]}
=Q_n^{(\text{I})}+Q_n^{(\text{II})},
\end{eqnarray}
where (with $W_n^2=\Omega^2 n + \Gamma^2$)
\begin{eqnarray}
Q_n^{(\text{I})}&=&\Gamma
\frac{2\Gamma^2[\Omega^2-4\varepsilon^2(2n+1)]+[4\Gamma^2+\Omega^2(2n+1)]
\left[\varepsilon^2-W^2_n+\sqrt{(\varepsilon^2-W^2_n)^2+4\varepsilon^2\Gamma^2}\right]}
{2[(4 \varepsilon \Gamma)^2+\Omega^4]\sqrt{(\varepsilon^2-W^2_n)^2+4\varepsilon^2\Gamma^2}
\sqrt{\varepsilon^2-W^2_n+\sqrt{(\varepsilon^2-W^2_n)^2+4\varepsilon^2\Gamma^2}}},
\\
Q_n^{(\text{II})}&=&\Gamma
\frac{-2\Gamma^2[\Omega^2-4\varepsilon^2(2n+1)]+[4\Gamma^2+\Omega^2(2n+1)]
\left[\varepsilon^2-W^2_{n+1}+\sqrt{(\varepsilon^2-W^2_{n+1})^2+4\varepsilon^2\Gamma^2} \right]}
{2[(4 \varepsilon \Gamma)^2+\Omega^4]\sqrt{(\varepsilon^2-W^2_{n+1})^2+4\varepsilon^2\Gamma^2}
\sqrt{\varepsilon^2-W^2_{n+1}+\sqrt{(\varepsilon^2-W^2_{n+1})^2+4\varepsilon^2\Gamma^2}}}.
\nonumber
\\
\end{eqnarray}
Since the term $Q_n^{(\text{II})}$ is dominated by $\varepsilon\sim W_{n+1}$, it
is convenient to shift the summation over $n$ for this term.
For definiteness, we use the hard cut-off for the summation over Landau levels, such that the highest Landau level involved in the summation
is $(N_\text{max}-1)+1=N_\text{max}$ from $Q^{(\text{II})}$:
\begin{equation}
\sum_{n=0}^{N_\text{max}-1} Q_n = Q_0^{(\text{I})}+\sum_{n=1}^{N_\text{max}-1} \left[Q_n^{(\text{I})}+Q_{n-1}^{(\text{II})}\right]
+Q_{N_\text{max}-1}^{(\text{II})},
\label{sumQn}
\end{equation}
where
\begin{eqnarray}
Q_0^{(\text{I})}&=&\frac{\Gamma \Omega^2}{2[(4\varepsilon \Gamma)^2+\Omega^4]},
\label{Q0I}
\\
Q_{N_\text{max}-1}^{(\text{II})}&\simeq& \frac{4\varepsilon^2\Gamma^2 \sqrt{N_\text{max}}}{\Omega[(4\varepsilon \Gamma)^2+\Omega^4]},
\label{Qmax}
\end{eqnarray}
and we write
\begin{eqnarray}
Q_n^{(\text{I})}+Q_{n-1}^{(\text{II})}&=&\frac{\sqrt{2}\Gamma \varepsilon}{(4 \varepsilon \Gamma)^2+\Omega^4}\ q_n,
\\
q_n&=&q_{n,1}-q_{n,2}.
\end{eqnarray}
Here we have split $q_n$ into two parts related to the asymmetry of each Landau level:
\begin{eqnarray}
q_{n,1}&=& \Omega^2 n\frac{ \sqrt{\varepsilon^2-W^2_{n}+\sqrt{(\varepsilon^2-W^2_{n})^2+4\varepsilon^2\Gamma^2}}}{\sqrt{(\varepsilon^2-W^2_{n})^2+4\varepsilon^2\Gamma^2}},
\nonumber
\\
\label{qn1}
\end{eqnarray}
\begin{eqnarray}
q_{n,2}&=& \frac{\Gamma}{\varepsilon}\left[\varepsilon^2+W^2_{n}-\sqrt{(\varepsilon^2-W^2_{n})^2+4\varepsilon^2\Gamma^2}\right]
 \frac{ \sqrt{W^2_{n}-\varepsilon^2+\sqrt{(\varepsilon^2-W^2_{n})^2+4\varepsilon^2\Gamma^2}}}{\sqrt{(\varepsilon^2-W^2_{n})^2+4\varepsilon^2\Gamma^2}}.
\label{qn2}
\end{eqnarray}
\end{widetext}
For weak disorder, $\Gamma(\varepsilon)\ll \varepsilon$, and $q_{n,1}\gg q_{n,2}$.
We note that these functions have resonant structure and take their maximal values when
\begin{equation}
\varepsilon^2 \simeq \Omega^2 n +\frac{2}{\sqrt{3}}\Gamma\Omega\sqrt{n}.
\end{equation}
Comparing Eqs.~(\ref{qn1}) and (\ref{qn2}) at resonances, we see that the maximal value of $q_n$
is dominated by $q_{n,1}$:
\begin{equation}
q_{n,1}\simeq \frac{\varepsilon^{3/2}}{\sqrt{2}\Gamma^{1/2}},
\qquad q_{n,2}\simeq \sqrt{2}\Gamma^{1/2} \varepsilon^{1/2} \ll q_{n,1}.
\end{equation}
One can check that $q_{n,1}\gg q_{n,2}$ in the whole range of energies $W_{n}<\varepsilon<W_{n+1}$.
Therefore, in what follows we will disregard the contribution of $q_{n,2}$, using $q_n\simeq q_{n,1}$.
Comparing Eq.~(\ref{qn1}) and (\ref{Gamman}), we find [see also Eq.~(\ref{SCEq})]:
\begin{equation}
q_n \simeq \Omega^2 n \frac{\sqrt{2}\Gamma^{(n)}(\varepsilon)}{A \varepsilon}
= \Omega^2 n \sqrt{2}\left(\frac{\Gamma}{A \varepsilon}-\frac{2\varepsilon}{\Omega^2}\right)
\label{qn-gamman}
\end{equation}

Let us now consider energy $\varepsilon$ located between the Landau levels $N$ and $N+1$.
We split the sum over Landau levels in Eq.~(\ref{sumQn})
as follows:
\begin{eqnarray}
&&\sum_{n=1}^{N_\text{max}-1}\left[Q_n^{(\text{I})}+Q_{n-1}^{(\text{II})}\right]
=
\frac{\sqrt{2}\Gamma \varepsilon}{(4 \varepsilon \Gamma)^2+\Omega^4}\ \sum_{n=1}^{N_\text{max}-1} q_n
\nonumber
\\
&&=\frac{\sqrt{2}\Gamma \varepsilon}{(4 \varepsilon \Gamma)^2+\Omega^4}\
\left\{\sum_{n=1}^{N-1} q_n + q_N + q_{N+1} +  \sum_{n=N+2}^{N_\text{max}-1} q_n \right\}.
\nonumber
\\
\label{sum-n}
\end{eqnarray}

When the Landau-level broadening is smaller than the distance between the neighboring levels,
$W_N-W_{N+1}\sim \Omega^2/\varepsilon\gg \Gamma(\varepsilon)$, i.e., $\varepsilon\ll \varepsilon_{**}=\Omega(\Omega/A)^{1/3}$,
we can neglect $\Gamma$ in the contributions
of all Landau levels with $n<N$. Replacing the sums by integrals,
we find
\begin{eqnarray}
\sum_{n=1}^{N-1} q_n &\simeq& \sum_{n=1}^{N-1} \frac{\sqrt{2}\Omega^2 n}{\sqrt{\varepsilon^2-\Omega^2 n}}
\simeq \frac{4\sqrt{2}}{3}\frac{\varepsilon^3}{\Omega^2}.
\end{eqnarray}
This contribution to Eq.~(\ref{sumQn}) is by a factor $\varepsilon^4/\Omega^4$ larger than the $n=0$ term, Eq.~(\ref{Q0I}),
so the latter can be neglected.
For $n>N+1$ we expand $q_n$ in $\Gamma$ and get
\begin{eqnarray}
&&\sum_{n=N+2}^{N_\text{max}-1} q_n \simeq \sum_{n=N+2}^{N_\text{max}-1}
\frac{\sqrt{2} \Gamma(2\varepsilon^2-\Omega^2 n)}{(\Omega^2 n-\varepsilon^2)^{3/2}}
\nonumber
\\
&&
\simeq
-\frac{2\sqrt{2}\varepsilon \Gamma}{\Omega}\sqrt{N_\text{max}}
+\frac{2\sqrt{2}\varepsilon^3 \Gamma}{\Omega^2\sqrt{\Omega^2(N+2)-\varepsilon^2}}.
\end{eqnarray}
The term proportional to $\sqrt{N_\text{max}}$ exactly cancels the contribution of Eq.~(\ref{Qmax})
in the sum over $n$, Eq.~(\ref{sumQn}). The second term here is of the order of $\Gamma/\Omega$ [remember that $\Omega^2N<\varepsilon^2<\Omega^2(N+1)$]
and can be neglected for $\varepsilon<\Omega(\Omega/A)^{1/2}$ compared to the contribution
of $n<N$. Then Eq.~(\ref{sumQn}) takes the form
\begin{eqnarray}
\sum_{n=0} Q_n &\simeq&
 \frac{2\Gamma \varepsilon^4}{\Omega^2[(4 \varepsilon \Gamma)^2+\Omega^4]}
 \left[
\frac{4}{3}
 +
\frac{\Omega^2}{\sqrt{2}\varepsilon^3 }\left(q_N+q_{N+1}\right)\right].
\nonumber
\\
\label{3terms}
\end{eqnarray}
According to Eq.~(\ref{qn-gamman}), the second term in Eq.~(\ref{3terms}) is given by
\begin{equation}
\frac{\Omega^2}{\sqrt{2}\varepsilon^3 }\left(q_N+q_{N+1}\right)
\sim \frac{\Omega^2}{\varepsilon}\left(\frac{\Gamma(\varepsilon)}{A \varepsilon}-\frac{2\varepsilon}{\Omega^2}\right).
\end{equation}
This term may dominate the sum in Eq.~(\ref{3terms}) only for $\varepsilon<\varepsilon_*\sim \Omega(\Omega/A)^{1/5}$
and when $\varepsilon$ is located close to a Landau level center. We will return to this case later
and first analyze the opposite (simpler) regime of $\varepsilon>\varepsilon_*$.

For $\varepsilon>\varepsilon_{**}=\Omega(\Omega/A)^{1/3}$, the level broadening in $q_n$ is larger than the distance
$|\varepsilon-W_n|$ for Landau levels sufficiently close to $N$:
\begin{eqnarray}
&& \Gamma_N>|\varepsilon-W_n|\sim |W_N-W_n| \nonumber
\\
&& \Rightarrow \quad
|N-n|<N_\Gamma\equiv \frac{\sqrt{N}\Gamma_N}{\Omega}\sim \frac{A}{\Omega}\left(\frac{\varepsilon}{\Omega}\right)^3.
\end{eqnarray}
For such values of $n$ we can neglect $\varepsilon^2-W^2_{n}$ as compared to $\varepsilon\Gamma$
in Eq.~(\ref{qn1}):
\begin{equation}
q_{n}= \frac{\Omega^2 n}{\sqrt{2\varepsilon\Gamma}}.
\label{qn1-Gamma}
\end{equation}
In this case, the sum over $n$ in Eq.~(\ref{sum-n}) is written as
\begin{eqnarray}
\sum_{n=1}^{N_\text{max}-1} q_n
&=&\left\{\sum_{n=1}^{N-N_\Gamma-1}+\! \sum_{n=N-N_\Gamma}^{N+N_\Gamma}  +\!  \sum_{n=N+N_\Gamma+1}^{N_\text{max}-1}\right\} q_n,
\nonumber
\\
\sum_{n=1}^{N-N_\Gamma-1}\!\! q_n&\simeq& \frac{4\sqrt{2}}{3}\frac{\varepsilon^3}{\Omega^2}
\left[1-\frac{3}{2}\sqrt{ \frac{N_\Gamma}{N}}\right],
\\
\sum_{n=N-N_\Gamma}^{N+N_\Gamma}\!\!q_n&\simeq&N_\Gamma^2\frac{\Omega^2}{\sqrt{2\varepsilon\Gamma}}
=\frac{\varepsilon^{3/2}\Gamma^{3/2}}{\Omega^2},
\\
\sum_{n=N+N_\Gamma+1}^{N_\text{max}-1}\!\! q_n &\simeq& -\frac{2\sqrt{2}\varepsilon \Gamma}{\Omega}\sqrt{N_\text{max}}
+\frac{2\sqrt{2}\varepsilon^2 \Gamma}{\Omega^2}.
\label{sum-n-Gamma}
\end{eqnarray}
We see that for $\varepsilon>\varepsilon_{**}$ again the first term ($n<N-N_\Gamma$) dominates,
yielding the same result as for $\varepsilon<\varepsilon_{**}$ (clearly, $N_\Gamma\ll N$,
in view of $\Gamma(\varepsilon)\ll \varepsilon$).
Including the vertex correction calculated in Appendix \ref{app:vertcorr},
we arrive at Eqs.~\eqref{sumQn-result} and \eqref{sigmaxx-class} of the main text.

Let us now return to the case of lower temperatures, $\Omega<T<\varepsilon_*\sim \Omega(\Omega/A)^{1/5}$.
In this case, the contribution of the Landau level $N$ closest to the energy $\varepsilon$ in the sum in Eq.~(\ref{3terms})
should be analyzed. In order to estimate this contribution,
we replace the integral over energy in Eq.~(\ref{sigmaQ}) by a sum over regions of width $\Gamma_N$
around Landau levels,
use Eq.~(\ref{qn-gamman}), replace $\varepsilon$ by $W_N$, and replace $\Gamma^{(N)}(\varepsilon)$ there by its maximal value
$\Gamma^{(N)}(W_N)\equiv \Gamma_N \sim A^{2/3}\Omega^{1/3} N^{1/6}$.
As a result, we get
\begin{eqnarray}
\sigma_{xx}^{(N)}
&\sim& \frac{e^2\Omega^2}{A T v} \sum_{N<(T/\Omega)^2}
\Gamma_N\ \frac{\Gamma_N^2 W_N^2}{(4 W_N \Gamma_N)^2+\Omega^4}
\\
&\sim&
\frac{e^2 \gamma T^4}{\Omega^2 v^4}
\propto \frac{ \gamma T^4}{H}
\label{sigma-N}
\end{eqnarray}
In addition to this contribution,
there is a contribution of the tail at $\Gamma_N<\varepsilon-W_N<\Omega(\Omega/W_N)^3$, see Fig. \ref{fig3}
and Eq.~(\ref{right-tail}). The integral over $|\varepsilon-W_N|$ is logarithmic since $\Gamma^2(\varepsilon)$
decays as $(\varepsilon-W_N)^{-1}$ in this range and thus enhances the result (\ref{sigma-N}) by a logarithmic factor.
We see that the contribution to the conductivity of the Landau level $N$ is smaller than the semiclassical
contribution, Eq.~(\ref{sigmaxx-*-**}),
by factor $\Omega^2/T^2\ll 1$ (up to the logarithm) and can be neglected.

\end{document}